**Phonons of electronic crystals in two-dimensional semiconductor moiré patterns**

Hongyi Yu[1,2*], Jiyong Zhou[1]

[1] Guangdong Provincial Key Laboratory of Quantum Metrology and Sensing & School of Physics and Astronomy, Sun Yat-Sen University (Zhuhai Campus), Zhuhai 519082, China

[2] State Key Laboratory of Optoelectronic Materials and Technologies, Sun Yat-Sen University (Guangzhou Campus), Guangzhou 510275, China

* E-mail: yuhy33@mail.sysu.edu.cn

**Abstract:** We theoretically studied the phonon properties of the triangular-, stripe- and honeycomb-type electronic crystals recently found in two-dimensional semiconductor moiré patterns. By analyzing the phonon dispersions, we found the interaction induced lattice deformation in the zigzag-stripe crystal results in a much higher dynamical stability than in the linear-stripe crystal. Moreover, chiral phonons with finite magnetizations and large Berry curvatures can emerge in triangular and honeycomb crystals under time-reversal or inversion symmetry breaking. The small effective mass of the electrons allows the selective and efficient generation of chiral phonons from the optical activity of zone-center phonons combined with the anharmonicity, facilitating the realization of the phonon Hall effect. These findings point to an exciting new platform for exploring chiral phonons and the related topological phononic devices.

**I. Introduction**

The formation of long-wavelength moiré patterns in van der Waals stacking of layered two-dimensional (2D) materials has introduced a new platform for studying exotic quantum phenomena [1,2]. Since the discovery of Mott insulator states and superconductivity in twisted bilayer graphene [3,4], there has been a surge of research interest about the strong correlation physics in layered moiré materials, especially in the semiconducting transition metal dichalcogenides (TMDs). By continuously tuning the doping density in the bilayer TMDs moiré system, a variety of electronic crystals, including Mott insulators, generalized Wigner crystals and stripe crystals, have been detected at integer and fractional fillings [5-18]. Early research works for electronic crystals in semiconductor quantum wells usually rely on strong magnetic fields to quench the electron kinetic energy and enhance the localization, where the formed Wigner crystals have large wavelengths (~ 30 nm) and can only be in the triangular lattice [19]. Due to the long-range nature of the Coulomb interaction, the low-energy excitations of the electronic crystal are dominated by the collective vibration of the crystal sites (i.e., phonons), which is often viewed as a signature of the crystalline order and has been well studied for the triangular Wigner crystal [20-24]. On the other hand in TMDs moiré systems, the formation of the electronic crystal is a result of the enhanced Coulomb interaction in 2D geometry combined with the confinement from the moiré potential. A series of electronic crystals not restricted to the triangular lattice have been observed under zero magnetic field, with much shorter wavelengths (~ 10 nm or shorter) and stronger inter-site interactions. However, despite the rapid progress in the research of various electronic crystals in TMDs moiré patterns, a thorough investigation about their phonon modes have so far been lacking.

Here we investigate the phonon properties of these electronic crystals recently found in TMDs moiré systems, including the triangular, linear-stripe, zigzag-stripe and honeycomb crystals. The obtained phonon dispersions indicate that the moiré confinement is essential for stabilizing the stripe and honeycomb crystals. When the moiré confinement is weak, the zigzag-stripe crystal is found to have a much higher dynamical stability than the linear-stripe crystal, as in the former case the inter-site interaction can deform the lattice to resemble the most stable triangular crystal. Moreover, chiral phonons with finite magnetizations and large Berry curvatures can emerge in triangular and honeycomb crystals under time-reversal symmetry breaking, or in honeycomb crystals under inversion symmetry breaking [25]. These chiral phonons are located near the Brillouin zone center **Γ** or corners **K/K̄**, and have great tunability through external electric, magnetic and optical fields due to the small effective mass of the electrons as compared to their ionic counterparts. We find that the **Γ**-phonon of the electronic crystals can be directly generated by absorbing an infrared photon or through a two-photon Raman process. Meanwhile the anharmonicity from the moiré/interaction potential can efficiently convert a **Γ**-phonon into two phonons at **K** and **K̄**, respectively, allowing the optical generation of (**K**,**K̄**)-phonon pairs. The chirality and Berry curvatures of the generated phonons can be selected through the choices of the photon polarization and frequency. Under a driving force induced by the spatially inhomogeneous heterostrain and/or twist angle or interlayer bias, a Hall effect of the chiral phonons can emerge which can be detected from the accumulated magnetization and temperature gradient at the sample edge.

The rest of the paper is organized as follows. In Sec. II we show the calculated phonon dispersions of the triangular-, stripe- and honeycomb-type electronic crystals, and discuss the optical detection of the phonons at **Γ**-point. Sec. III focuses on the chiral phonons in time-reversal-asymmetric or inversion-asymmetric triangular and honeycomb crystals, and the corresponding phonon Hall effect. In Sec. IV we give a quantitative analysis to the effect of anharmonicities. Sec. V is the discussion.

**II. Phonon dispersions of various electronic crystals**

Throughout the paper we set $e = \hbar = \epsilon_0 = 1$, with $e$ the charge of an electron, $\hbar$ the reduced Planck constant and $\epsilon_0$ the vacuum permittivity. For the zigzag-stripe and honeycomb crystals where each unit cell contains two sublattice sites (denoted as $\mu = A, B$), the system Hamiltonian is

$$\hat{H}_{\text{phonon}} = \sum_\mu \sum_n \left( -\frac{1}{2m_\mu} \frac{\partial^2}{\partial r_{\mu,n}^2} + V_{\text{moiré}}(\mathbf{R}_{\mu,n} + \mathbf{r}_{\mu,n}) \right)$$
$$+ \frac{1}{2} \sum_{\mu\nu} \sum_{n'n} V(\mathbf{R}_{\mu,n'} - \mathbf{R}_{\nu,n} + \mathbf{r}_{\mu,n'} - \mathbf{r}_{\nu,n}). \quad (1)$$

On the above right-hand-side, the first line is the single-particle Hamiltonian including the kinetic energy and the moiré potential, and the second one is the interaction potential where the summation is over all interacting pairs with $\mathbf{R}_{\mu,n'} - \mathbf{R}_{\nu,n} \neq \mathbf{0}$. Here we assume the system is deep in the crystal regime, and write the sublattice equilibrium positions in $n$-th unit cell as $\mathbf{R}_{A,n} = \mathbf{R}_n$ and $\mathbf{R}_{B,n} = \mathbf{R}_n - \mathbf{R}_{AB}$, respectively. $\mathbf{r}_{\mu,n}$ is the variable which describes the motion of the $\mu = A, B$ sublattice in $n$-th unit cell, whose magnitude is assumed to be much smaller than the wavelength $\lambda$ of the moiré pattern. In some cases the electron/hole effective masses $m_A$ and $m_B$ at the two sublattices are different, however they can be made equivalent by replacing $\mathbf{r}_{B,n}$ with $\mathbf{r}'_{B,n}\sqrt{m_A/m_B}$. In this work we set $m_A = m_B = m$. For the triangular and linear-stripe crystals where each unit cell contains only one site, the phonon Hamiltonian can be obtained by simply dropping all the $B$ sublattice related terms.

The periodic moiré potential can be generally written in the form [26-29]

$$V_{\text{moiré}}(\mathbf{r}) \approx \Delta_1 \left| 1 + e^{i(\mathbf{b}_1 \cdot \mathbf{r} - \frac{2\pi}{3})} + e^{-i(\mathbf{b}_2 \cdot \mathbf{r} - \frac{2\pi}{3})} \right|^2$$
$$+ \Delta_2 \left| 1 + e^{i(\mathbf{b}_1 \cdot \mathbf{r} + \frac{2\pi}{3})} + e^{-i(\mathbf{b}_2 \cdot \mathbf{r} + \frac{2\pi}{3})} \right|^2. \quad (2)$$

Here $\mathbf{b}_1$ and $\mathbf{b}_2$ are the primitive reciprocal lattice vectors of the triangular moiré pattern, which is different from those of the electronic crystal when the filling factor $\nu < 1$. The landscape of the moiré potential is determined by the positive parameters $\Delta_1, \Delta_2$ that are on the order of 1-10 meV from numerical calculations [26-28]. In the vicinity of the potential minimum located at $\mathbf{r} = \mathbf{0}$, the moiré potential can be approximated by a harmonic form $V_{\text{moiré}}(\mathbf{r}) \approx \frac{\gamma}{2\lambda^2} \mathbf{r}^2$ with $\gamma \equiv 8\pi^2(\Delta_1 + \Delta_2)$ the moiré confinement strength. We estimate $\gamma$ to be in the range of 0.1-1 eV. The moiré potential can also form a honeycomb lattice when $\Delta_2 \ll \Delta_1$ [26,30-37], where the two local minima generally have different confinement strengths and their difference can be tuned by an interlayer bias [34]. However in the case with a filling factor $\nu \leq 1$, only the global minima will be occupied.

The Coulomb interaction $V(\mathbf{r})$ is modified by the atomically-thin geometry of the layered structure, which can be expressed in the Rytova-Keldysh form [38-40]

$$V(\mathbf{r}) = \frac{\pi}{2\epsilon r_0}\left[H_0\left(\frac{r}{r_0}\right) - Y_0\left(\frac{r}{r_0}\right)\right], \qquad (3)$$

where $H_0$ and $Y_0$ are the Struve and 2nd-kind Bessel functions, respectively. $r_0 = 2\pi\alpha_{2D}/\epsilon$ is the 2D screening length, with $\alpha_{2D}$ the 2D polarizability of the layered material and $\epsilon$ the relative dielectric constant of the environment. For a fixed $\epsilon$, larger $r_0$ implies stronger 2D screening and weaker Coulomb interaction strength. Different monolayer TMDs have similar screening lengths $r_0 \approx 5/\epsilon$ nm [41], while $r_0 \approx 10/\epsilon$ nm can approximately describe Coulomb interactions in bilayer TMDs. In the limit $r_0 \to 0$, the interaction returns to the traditional form in 3D homogeneous space, i.e., $V(\mathbf{r}) \to \frac{1}{\epsilon r}$.

We expand the moiré and interaction potentials up to the 2nd-order of $\mathbf{r}_{\mu,n} \equiv (x_{\mu,n}, y_{\mu,n})$, i.e., the harmonic approximation (see Appendix A for a discussion about the mean-square-displacement of the electron and the validity of the harmonic approximation). To facilitate the analysis on the phonon chirality, below we adopt the left- and right-handed basis and introduce $r_{\mu,\mathbf{k}}^{\pm} \equiv \frac{1}{\sqrt{N}}\sum_n \frac{x_{\mu,n} \pm i y_{\mu,n}}{\sqrt{2}} e^{-i\mathbf{k}\cdot\mathbf{R}_{\mu,n}}$ with $N$ the total number of the crystal unit cells. The phonon Hamiltonian is then in a form $\hat{H}_{\text{phonon}} = E_0 + \sum_{\mathbf{k}}\hat{H}_{\mathbf{k}}$ with $E_0 \equiv \sum_{\mu n} V_{\text{moiré}}(\mathbf{R}_{\mu,n}) + \frac{1}{2}\sum_{\mu\nu}\sum_{n'n} V(\mathbf{R}_{\mu,n'} - \mathbf{R}_{\nu,n})$ a constant potential, and

$$\hat{H}_{\mathbf{k}} = -\frac{1}{2m}\left(\frac{\partial^2}{\partial r_{A,\mathbf{k}}^+ \partial r_{A,-\mathbf{k}}^-} + \frac{\partial^2}{\partial r_{A,\mathbf{k}}^- \partial r_{A,-\mathbf{k}}^+}\right) - \frac{1}{2m}\left(\frac{\partial^2}{\partial r_{B,\mathbf{k}}^+ \partial r_{B,-\mathbf{k}}^-} + \frac{\partial^2}{\partial r_{B,\mathbf{k}}^- \partial r_{B,-\mathbf{k}}^+}\right) + \frac{1}{2}\vec{r}_{\mathbf{k}}\vec{\mathbf{g}}_{\mathbf{k}}\vec{r}_{\mathbf{k}}^\dagger,$$

$$\vec{\mathbf{g}}_{\mathbf{k}} = \begin{pmatrix} g(\mathbf{k}) + \frac{\gamma_A}{\lambda^2} & g_{+-}(\mathbf{k}) & g^{AB}(\mathbf{k}) & g_{+-}^{AB}(\mathbf{k}) \\ g_{-+}(\mathbf{k}) & g(\mathbf{k}) + \frac{\gamma_A}{\lambda^2} & g_{-+}^{AB}(\mathbf{k}) & g^{AB}(\mathbf{k}) \\ g^{BA}(\mathbf{k}) & g_{+-}^{BA}(\mathbf{k}) & g(\mathbf{k}) + \frac{\gamma_B}{\lambda^2} & g_{+-}(\mathbf{k}) \\ g_{-+}^{BA}(\mathbf{k}) & g^{BA}(\mathbf{k}) & g_{-+}(\mathbf{k}) & g(\mathbf{k}) + \frac{\gamma_B}{\lambda^2} \end{pmatrix}. \qquad (4)$$

Here $\vec{r}_{\mathbf{k}} \equiv (r_{A,\mathbf{k}}^+, r_{A,\mathbf{k}}^-, r_{B,\mathbf{k}}^+, r_{B,\mathbf{k}}^-)$, $\frac{\gamma_\mu}{\lambda^2} = \nabla^2 V_{\text{moiré}}(\mathbf{R}_{\mu,n} + \mathbf{r})\big|_{\mathbf{r}=0}$. $\vec{\mathbf{g}}_{\mathbf{k}}$ is the dynamical matrix whose elements have the following expressions (see Appendix B for calculation details):

$$g(\mathbf{k}) = \frac{1}{2}\sum_{n\neq 0}(1 - \cos(\mathbf{k}\cdot\mathbf{R}_n))\nabla^2 V(\mathbf{R}_n + \mathbf{r})\big|_{\mathbf{r}=0} + \frac{1}{2}\sum_n \nabla^2 V(\mathbf{R}_n + \mathbf{r})\big|_{\mathbf{r}=\mathbf{R}_{AB}},$$

$$g_{\pm\mp}(\mathbf{k}) = \frac{1}{2}\sum_{n\neq 0}(1 - \cos(\mathbf{k}\cdot\mathbf{R}_n))\partial_\mp^2 V(\mathbf{R}_n + \mathbf{r})\big|_{\mathbf{r}=0} + \frac{1}{2}\sum_n \partial_\mp^2 V(\mathbf{R}_n + \mathbf{r})\big|_{\mathbf{r}=\mathbf{R}_{AB}},$$

$$g^{AB}(\mathbf{k}) = \left(g^{BA}(\mathbf{k})\right)^* = -\frac{1}{2}\sum_n e^{i\mathbf{k}\cdot(\mathbf{R}_n + \mathbf{R}_{AB})}\nabla^2 V(\mathbf{R}_n + \mathbf{r})\big|_{\mathbf{r}=\mathbf{R}_{AB}},$$

$$g_{\pm\mp}^{AB}(\mathbf{k}) = \left(g_{\mp\pm}^{BA}(\mathbf{k})\right)^* = -\frac{1}{2}\sum_n e^{i\mathbf{k}\cdot(\mathbf{R}_n+\mathbf{R}_{AB})}\partial_\mp^2 V(\mathbf{R}_n+\mathbf{r})\big|_{\mathbf{r}=\mathbf{R}_{AB}}.$$

Here we have used the notation $\partial_\pm \equiv \frac{\partial}{\partial x} \pm i\frac{\partial}{\partial y}$.

By diagonalizing the dynamical matrix $\overleftrightarrow{\mathbf{g}}_\mathbf{k}$, we obtain the normal coordinates $Q_{l,\mathbf{k}} = A_{l,\mathbf{k}}^+ r_{A,\mathbf{k}}^+ + A_{l,\mathbf{k}}^- r_{A,\mathbf{k}}^- + B_{l,\mathbf{k}}^+ r_{B,\mathbf{k}}^+ + B_{l,\mathbf{k}}^- r_{B,\mathbf{k}}^-$ and the phonon creation and annihilation operators $\hat{a}_{l,\mathbf{k}}^\dagger = \sqrt{\frac{m\omega_{l,\mathbf{k}}}{2}} Q_{l,-\mathbf{k}} - \frac{1}{\sqrt{2m\omega_{l,\mathbf{k}}}}\frac{\partial}{\partial Q_{l,\mathbf{k}}}$, $\hat{a}_{l,\mathbf{k}} = \sqrt{\frac{m\omega_{l,\mathbf{k}}}{2}} Q_{l,\mathbf{k}} + \frac{1}{\sqrt{2m\omega_{l,\mathbf{k}}}}\frac{\partial}{\partial Q_{l,-\mathbf{k}}}$. The phonon Hamiltonian then becomes $\hat{H}_\mathbf{k} = \sum_{l=1}^4 \omega_{l,\mathbf{k}}(\hat{a}_{l,\mathbf{k}}^\dagger \hat{a}_{l,\mathbf{k}} + \frac{1}{2})$. Here $\omega_{l,\mathbf{k}}$ is the phonon frequency with $l = 1$ to 4 denoting the index of the phonon branch from low- to high-frequencies. The four-component unit vector $(A_{l,\mathbf{k}}^+, A_{l,\mathbf{k}}^-, B_{l,\mathbf{k}}^+, B_{l,\mathbf{k}}^-)^T$, being the eigenvector of $\overleftrightarrow{\mathbf{g}}_\mathbf{k}$, describes the circular polarization of $l$-th phonon branch at $\mathbf{k}$. Here $A_{l,\mathbf{k}}^\pm$ and $B_{l,\mathbf{k}}^\pm$ correspond to the $\sigma^\pm$ component for the vibration of $A$ and $B$ sublattices, respectively.

### A. Triangular crystals

The triangular-type electronic crystal has been observed in monolayer $MoSe_2$ under a low doping density [17,18] and in $WS_2/WSe_2$ heterobilayer moiré superlattices at filling 1 or 1/3 [5-15]. Fig. 1(a) illustrates a typical triangular crystal formed in a moiré potential at filling 1. The corresponding phonon dispersions are given in Fig. 1(b-d). Note that we use $\mathbf{\Gamma}$, $\mathbf{K}$ and $\mathbf{M}$ to denote the high-symmetry points in the Brillouin zone of the electronic crystal, which are distinct from $\mathbf{\Gamma}_{TMD}$, $\mathbf{K}_{TMD}$ and $\mathbf{M}_{TMD}$ of the monolayer TMDs lattice. Here we set the electron/hole effective mass as $m \sim 0.5m_0$ (corresponding to electrons/holes in $\mathbf{K}_{TMD}$-valley) with $m_0$ the free electron mass. Fig. 1(b) shows the phonon dispersion for the triangular crystal with a wavelength $\lambda = 10$ nm without the moiré potential ($\gamma = 0$), under a weak environmental screening ($\epsilon = 1$) and three different values of $r_0$. The dispersion curve under $r_0 = 0$ is the same as that in early literatures [20]. The obtained phonon frequencies are real in the whole Brillouin zone, indicating dynamical stability of the crystal. In fact, without the moiré potential the triangular lattice is the only stable form in all 2D electronic crystals. The frequency values reach ~ 10 meV, which can further increase with the weakening of the screening or decreasing of the moiré wavelength. Compared to the ionic crystals, the electronic crystal sites have smaller effective masses but weaker interaction strengths, resulting in the same order of magnitudes for their phonon frequencies.

In the long-wavelength limit, the frequencies of the higher-energy longitudinal and lower-energy transverse modes scale as

$$\omega_{L,\mathbf{k}\to 0} = \sqrt{ak + bk^2}, \qquad \omega_{T,\mathbf{k}\to 0} = ck. \qquad (5)$$

Here $a = \frac{2\pi\rho}{m\epsilon}$ is a constant independent on the screening length $r_0$, with $\rho = \frac{2}{\sqrt{3}\lambda^2}$ the density of the electron/hole. This originates from the fact that the long-wavelength longitudinal mode leads to macroscopic charge modulations analogous to the 2D plasmon oscillation. The corresponding phonon frequency is then determined by $V_C(\mathbf{r})|_{r\gg r_0} \approx \frac{1}{\epsilon r}$ which is independent on $r_0$. The parameters $b$ and $c$ vary with $r_0$ as shown in Fig. 1(c), and $c \sim 0.1$ μm/ps corresponds to the group velocity of the transverse mode.

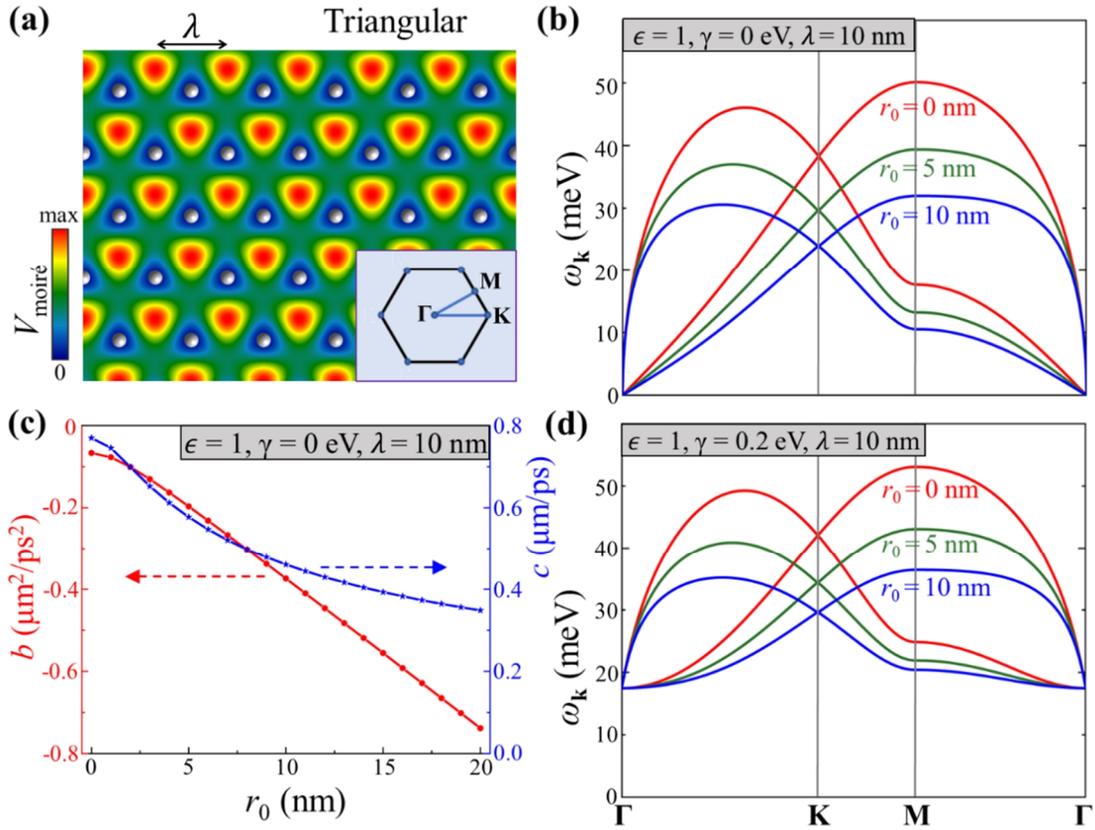

**Figure 1. The phonon dispersions of triangular crystals.** (a) Schematic illustration of a triangular moiré potential, and the formed electronic crystal at a filling factor of 1. Each circle corresponds to an electron/hole trapped at the moiré potential minimum. The inset shows the Brillouin zone. (b) The calculated phonon dispersions along the high-symmetry points **Γ-K-M-Γ**, for an electronic crystal without the moiré confinement ($\gamma = 0$). The other parameters are indicated in the figure. (c) The $b$ and $c$ values as functions of $r_0$. Here $b$ is related to the dispersion of the longitudinal branch in the long wavelength limit, and $c$ is the group velocity of the transverse branch in the long wavelength limit (see Eq. (5)). (d) The phonon dispersions under a finite moiré confinement strength $\gamma = 0.2$ eV.

When a moiré potential with a finite confinement strength $\gamma$ is introduced, the phonon becomes the mixture of the traditional phonon from the inter-site interaction and the harmonic oscillator mode in the moiré trap, with a nonzero frequency $\omega_{\mathbf{k}}(\gamma) = \sqrt{(\omega_{\mathbf{k}}(\gamma=0))^2 + \frac{\gamma}{m\lambda^2}}$ (Fig. 1(d)). The two degenerate phonons at $\boldsymbol{\Gamma}$-point correspond to acoustic modes with all sites oscillate in-phase. Hereafter we use $\boldsymbol{\Gamma}_{A+}$ and $\boldsymbol{\Gamma}_{A-}$ to denote the two $\boldsymbol{\Gamma}$-phonons with polarization vectors $(A^+_{1,\boldsymbol{\Gamma}}, A^-_{1,\boldsymbol{\Gamma}}) = \mathbf{(1,0)}$ and $(A^+_{2,\boldsymbol{\Gamma}}, A^-_{2,\boldsymbol{\Gamma}}) = \mathbf{(0,1)}$, respectively. Their frequency $\omega_{\boldsymbol{\Gamma}}(\gamma) = \lambda^{-1}\sqrt{\gamma/m}$ corresponds to the trapping frequency of the moiré potential and is independent on the interaction strength. The $\boldsymbol{\Gamma}_{A\pm}$-mode can then be excited by absorbing a $\sigma^\pm$ circularly polarized photon, analogous to excite the *s*-wave ground state in a harmonic trap to the *p*-wave excited state. When slightly away from $\boldsymbol{\Gamma}$, the longitudinal mode has a linear dispersion and can propagate in the electronic crystal with a group velocity $\frac{\pi\rho\lambda}{\epsilon\sqrt{m\gamma}} \sim 0.1$ µm/ps, whereas the transverse mode has a parabolic dispersion with the vanishing group velocity (see Fig. 1(d)).

### B. Stripe crystals

In heterobilayer $WS_2$/$WSe_2$ moiré superlattices, electronic linear- and zigzag-stripe crystals have been observed at 1/2 filling [10-13,16]. Fig. 2(a) schematically shows the linear-stripe crystal and its Brillouin zone, and Fig. 2(b,c) are the corresponding phonon dispersions. Unlike the triangular crystal, when $\gamma = 0$ the lower phonon branch of the linear-stripe crystal exhibits imaginary frequencies in a large region of the Brillouin zone, implying dynamical instability (Fig. 2(b)). A strong moiré trapping combined with a weak interaction strength can stabilize the crystal, such that all phonon frequencies become real. In this case the lowest frequency is located at the high-symmetry point $\mathbf{X}$. With the enhancement of the interaction strength by weakening the screening, the lowest frequency value decreases, which eventually crosses zero and becomes imaginary (see Fig. 2(c)).

The electronic crystal formed at 1/2 filling can also be in the zigzag-stripe form, as illustrated in Fig. 2(d). In this lattice configuration the overall Coulomb force felt by each crystal site is nonzero, which needs to be balanced by a displacement $\pm\delta x$ away from the moiré potential minimum [32]. The magnitude of the displacement falls in the range from 0 to $\lambda/4$. Fig. 2(d) inset shows the calculated value of $\delta x$ as a function of the wavelength $\lambda$, which is generally on the order of 1 nm. As analyzed below, this continuously tunable site-displacement plays an important role for the dynamical stability of the zigzag-stripe crystal.

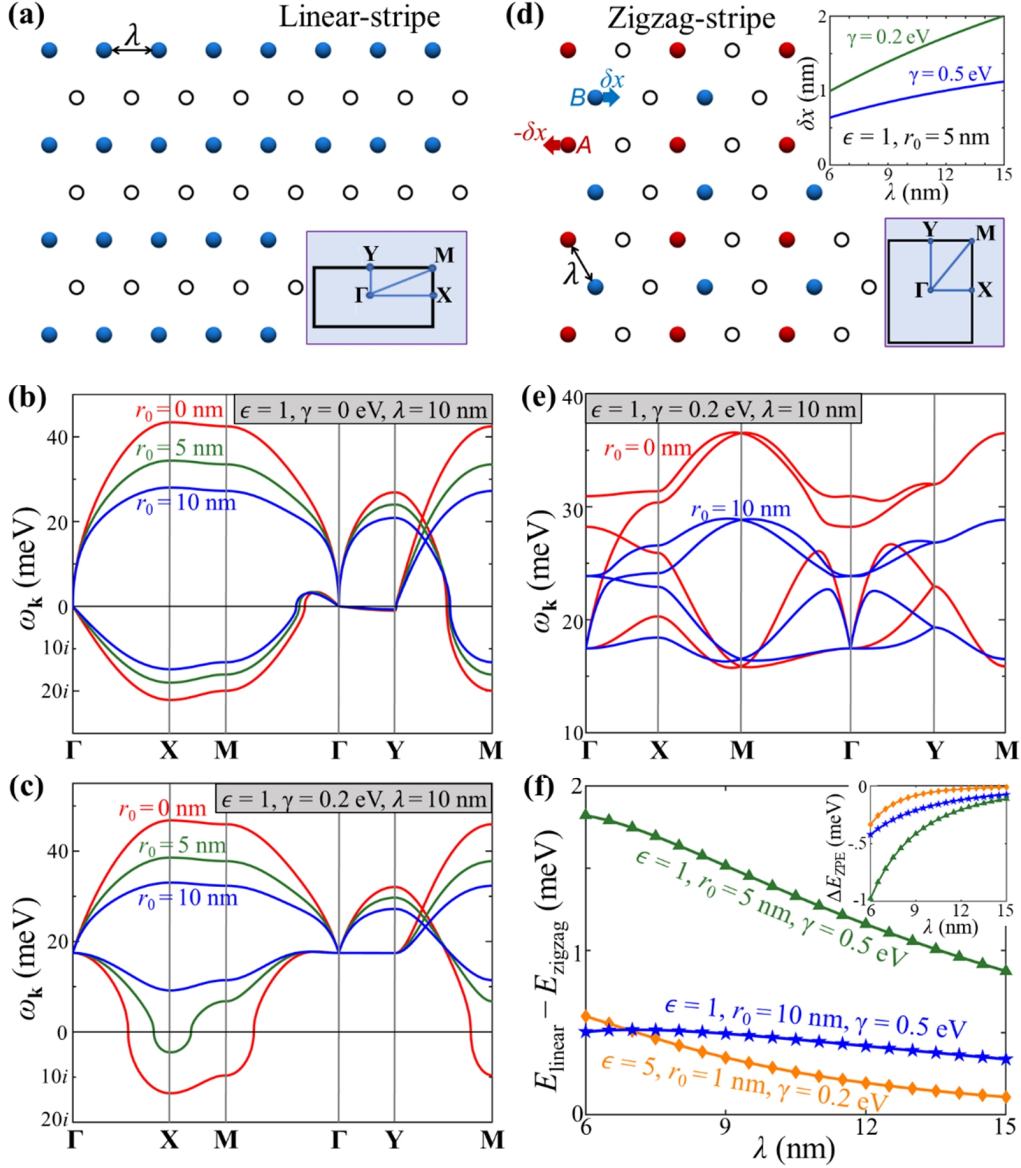

**Figure 2. The phonon dispersions of stripe crystals.** (a) Schematic illustration of the linear-stripe crystal formed in a triangular moiré superlattice at 1/2 filling, and its Brillouin zone. (b-c) The calculated phonon dispersions of the linear-stripe crystal along the high-symmetry points **Γ-X-M-Γ-Y-M** under (b) $\gamma = 0$ and (c) $\gamma = 0.2$ eV. The values of other parameters are indicated in the figures. The imaginary phonon frequencies imply that the corresponding crystals are dynamically unstable. (d) Schematic illustration of the zigzag-stripe crystal. Due to the unbalanced Coulomb forces, the two sublattices $A$ and $B$ are displaced from the moiré potential minima by $-\delta x$ and $\delta x$, respectively. The upper-right inset shows the magnitude of $\delta x$ as a function of $\lambda$, and the lower-right inset is the Brillouin zone. (e) The phonon dispersions of the zigzag-stripe crystals. (f) The ground state energy differences between the linear- and zigzag-stripe crystals under three sets of parameters. The inset shows the corresponding zero-point-energy (ZPE) differences.

In the absence of the moiré potential, both the linear- and zigzag-stripe crystals are dynamically unstable. However in contrast to Fig. 2(c) for the linear-stripe crystal, a tiny moiré confinement strength can already stabilize the zigzag-stripe crystal, implying that when the moiré potential is weak only the zigzag-stripe crystal can form at 1/2 filling (see Appendix C). Fig. 2(e) shows the corresponding phonon dispersions. We can see that when the screening length $r_0$ is reduced from 10 to 0 nm, the highest phonon frequency increases as expected due to the enhanced interaction strength. However, the lowest frequency near **M** barely changes with $r_0$, which is in sharp contrast to that of the linear-stripe crystal shown in Fig. 2(c). We find that the site-displacement $\delta x$ is essential for this anomalous phenomenon: when tuning $\delta x$ from 0 to $\lambda/4$, the deformed zigzag-stripe crystal becomes increasingly similar to a triangular crystal which greatly enhances its dynamical stability. In fact, when artificially setting $\delta x = 0$ the lowest phonon frequency of the zigzag-stripe crystal becomes sensitive to $r_0$ just like the linear-stripe case, and its stability under a weak moiré potential disappears (see Appendix C).

We have also compared the ground state energies of the two stripe crystals. The ground state energy per site for the zigzag-stripe crystal is

$$E_{\text{zigzag}} = \frac{\gamma}{2\lambda^2}\delta x^2 + \frac{1}{2}\left(\sum_{n\neq 0} V(\mathbf{R}_n) + \sum_n V(\mathbf{R}_n + \mathbf{R}_{AB})\right) + E_{\text{ZPE}}^{(\text{zigzag})}. \tag{6}$$

On the above right-hand-side, the first term is the moiré potential at the displaced equilibrium position, the second term is the static interaction potential, and the third term $E_{\text{ZPE}}^{(\text{zigzag})}$ is the contribution from the phonon zero-point-energy. That for the linear-stripe crystal can be obtained similarly, and the energy differences between the two stripe crystals are shown in Fig. 2(f). Although the linear-stripe crystal has a lower zero-point-energy than the zigzag-stripe crystal, the total ground state energy is lower for the zigzag-stripe crystal. This originates from the fact that the site-displacement increases the average distance between *A* and *B* sublattices, thus reduces the interaction potential. If artificially setting $\delta x = 0$, it is the linear-stripe crystal that has a lower total energy. The magnitude of $E_{\text{linear}} - E_{\text{zigzag}}$, however, is tiny (~ 1 meV), so generally both stripe crystals can be observed in experiments as long as they are stable.

### C. Honeycomb crystals with inversion symmetry

An inversion-symmetric honeycomb crystal (with $m_A = m_B = m$ and $\gamma_A = \gamma_B = \gamma$ for the two sublattices) can form either in triangular moiré superlattices at 2/3 filling (Fig. 3(a)) [6,10-13], or in homobilayers with honeycomb-type moiré patterns at a filling factor of 2 [30-36]. The corresponding phonon dispersions are shown in Fig. 3(b,c). Just like the stripe crystals, the honeycomb crystal is dynamically unstable without the moiré confinement (Fig. 3(b)). When a large enough $\gamma$ stabilizes the crystal, both the lowest and highest frequencies are located at the Brillouin zone corners **K**/**K̄**, see Fig. 3(c). Near **K**, the dynamical matrix can be expanded as

$$\vec{\mathbf{g}}_{\mathbf{K}+\mathbf{k}} = g(\mathbf{K}) + \begin{pmatrix} \gamma_A/\lambda^2 & \eta_{\mathbf{K}} k_+ & \zeta_{\mathbf{K}} k_- & g_{+-}^{AB}(\mathbf{K}) \\ \eta_{\mathbf{K}}^* k_- & \gamma_A/\lambda^2 & \eta_{\mathbf{K}}^{AB} k_+ & \zeta_{\mathbf{K}} k_- \\ \zeta_{\mathbf{K}}^* k_+ & \eta_{\mathbf{K}}^{AB*} k_- & \gamma_B/\lambda^2 & \eta_{\mathbf{K}} k_+ \\ g_{-+}^{BA}(\mathbf{K}) & \zeta_{\mathbf{K}}^* k_+ & \eta_{\mathbf{K}}^* k_- & \gamma_B/\lambda^2 \end{pmatrix} + O(k^2). \qquad (7)$$

Here $k_\pm \equiv k_x \pm i k_y$, $\eta_{\mathbf{K}} \equiv \left.\frac{\partial g_{+-}(\mathbf{k})}{\partial k_+}\right|_{\mathbf{k}=\mathbf{K}}$, $\zeta_{\mathbf{K}} \equiv \left.\frac{\partial g^{AB}(\mathbf{k})}{\partial k_-}\right|_{\mathbf{k}=\mathbf{K}}$ and $\eta_{\mathbf{K}}^{AB} \equiv \left.\frac{\partial g_{-+}^{AB}(\mathbf{k})}{\partial k_+}\right|_{\mathbf{k}=\mathbf{K}}$. The polarization vectors and the sublattice motions of the four phonon modes at $\mathbf{K}$ are summarized in Fig. 3(d), those at $\overline{\mathbf{K}}$ can be obtained by a time reversal. The polarization vectors of the lowest- and highest-frequency modes are $(A_{1,\mathbf{K}}^+, A_{1,\mathbf{K}}^-, B_{1,\mathbf{K}}^+, B_{1,\mathbf{K}}^-) = \left(\frac{1}{\sqrt{2}}, 0, 0, \frac{1}{\sqrt{2}}\right)$ and $(A_{4,\mathbf{K}}^+, A_{4,\mathbf{K}}^-, B_{4,\mathbf{K}}^+, B_{4,\mathbf{K}}^-) = \left(\frac{1}{\sqrt{2}}, 0, 0, \frac{-1}{\sqrt{2}}\right)$, respectively, which means the vibrations of the two sublattices have the same amplitude but are $\sigma^+$ and $\sigma^-$ circularly polarized for $A$ and $B$, respectively. Meanwhile the two degenerate intermediate modes at $\mathbf{K}$ have polarization vectors $(A_{2,\mathbf{K}}^+, A_{2,\mathbf{K}}^-, B_{2,\mathbf{K}}^+, B_{2,\mathbf{K}}^-) = (0,1,0,0)$ and $(A_{3,\mathbf{K}}^+, A_{3,\mathbf{K}}^-, B_{3,\mathbf{K}}^+, B_{3,\mathbf{K}}^-) = (0,0,1,0)$, respectively, implying that the vibration of one sublattice is circularly polarized whereas the other is still. In the vicinity of $\mathbf{K}$, the frequencies of the two intermediate modes are well separated from the lowest- and highest-frequency ones and can be well described by a 2x2 massless Dirac model.

Around $\mathbf{\Gamma}$ the dynamical matrix can be expanded as

$$\vec{\mathbf{g}}_{\mathbf{k}} = g(\mathbf{0}) + \begin{pmatrix} \gamma_A/\lambda^2 + \frac{ak}{2} & \frac{ak_-^2}{2k} & g^{AB}(\mathbf{0}) + \frac{ak}{2} & \eta_{\Gamma} k_+ + \frac{ak_-^2}{2k} \\ \frac{ak_+^2}{2k} & \gamma_A/\lambda^2 + \frac{ak}{2} & -\eta_{\Gamma}^* k_- + \frac{ak_+^2}{2k} & g^{AB}(\mathbf{0}) + \frac{ak}{2} \\ g^{BA}(\mathbf{0}) + \frac{ak}{2} & -\eta_{\Gamma} k_+ + \frac{ak_-^2}{2k} & \gamma_B/\lambda^2 + \frac{ak}{2} & \frac{ak_-^2}{2k} \\ \eta_{\Gamma}^* k_- + \frac{ak_+^2}{2k} & g^{BA}(\mathbf{0}) + \frac{ak}{2} & \frac{ak_+^2}{2k} & \gamma_B/\lambda^2 + \frac{ak}{2} \end{pmatrix} + O(k^2). \qquad (8)$$

Here $\eta_{\Gamma} \equiv \left.\frac{\partial (g_{+-}^{AB}(\mathbf{k}) - g_{+-}(\mathbf{k}))}{\partial k_+}\right|_{\mathbf{k}=0}$ and $a = \frac{2\pi\rho}{m\epsilon}$. At $\Gamma$-point, both the low-frequency acoustic and high-frequency optical modes are doubly degenerate. The acoustic modes $\Gamma_{A\pm}$ correspond to all sites vibrate with the same amplitude and same phase, whereas the optical modes $\Gamma_{O\pm}$ correspond to $A$ and $B$ sublattices vibrate with the same amplitude but has a phase difference of $\pi$.

In such an inversion-symmetric honeycomb crystal, the vibration of a given sublattice can be circularly polarized (see Fig. 3(d)). However, the parity-time symmetry ensures the average circular polarization or chirality vanishes for every non-degenerate mode. The emergence of chiral phonons requires either time-reversal or inversion symmetry breaking [25], which we will discuss in detail in Sec III below.

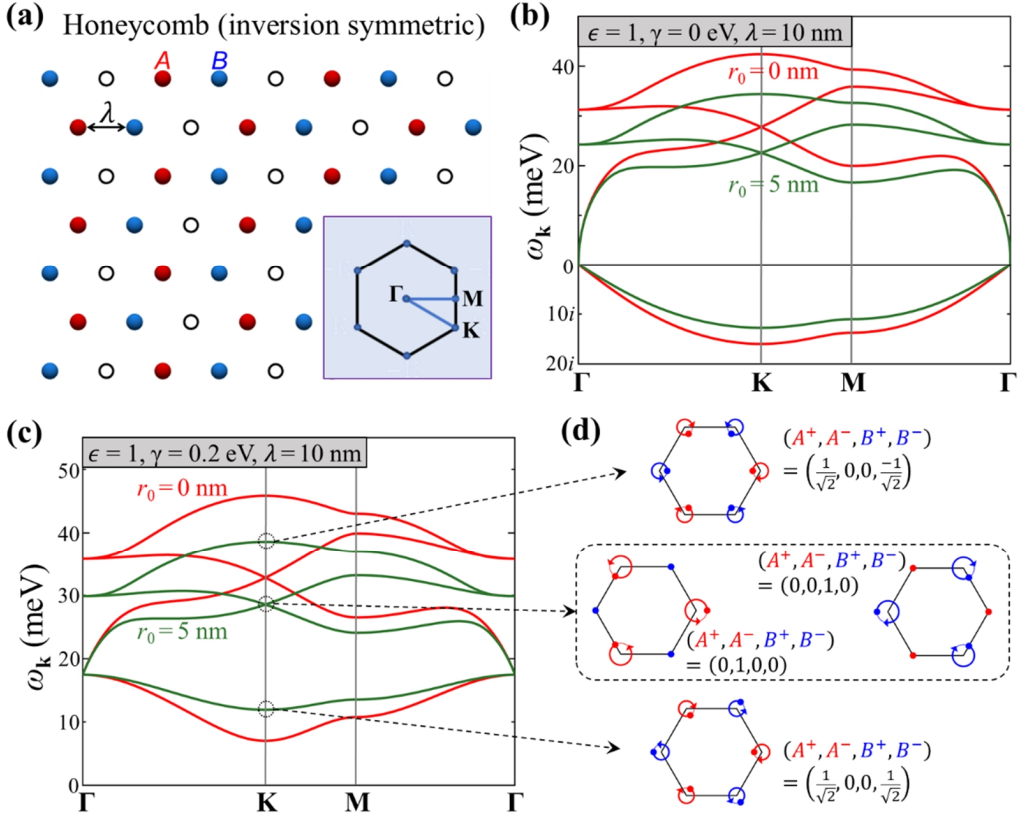

**Figure 3. The phonon dispersions of inversion-symmetric honeycomb crystals.** (a) Schematic illustration of an inversion-symmetric honeycomb crystal formed in a triangular moiré superlattice at 2/3 filling, and its Brillouin zone. (b-c) The calculated phonon dispersions along the high-symmetry points **Γ**-**K**-**M**-**Γ**, under the moiré confinement strengths of (b) $\gamma = 0$ and (c) $\gamma = 0.2$ eV. (d) The polarization vectors and vibration motions of $A$ and $B$ sublattices for the **K**-point phonon modes.

### D. Optical properties of phonons at Γ

The phonons at **Γ** can contribute to the infrared absorption and/or Raman scattering, thus plays an important role in optical properties of electronic crystals. Under the harmonic approximation, all the above discussed electronic crystals are inversion-symmetric and a **Γ**-phonon has either the even or odd parity. The acoustic modes $\mathbf{\Gamma_{A\pm}}$ have odd parities thus are expected to be infrared-active but Raman-inactive. Whereas the optical modes $\mathbf{\Gamma_O}$ in zigzag-stripe and honeycomb crystals have even parities thus are Raman-active but infrared-inactive. However when going beyond the harmonic approximation, the cubic expansion term from the moiré potential violates the inversion symmetry. Such an anharmonicity gives rise to three-phonon processes which can activate the $\mathbf{\Gamma_{A\pm}}$-involved Raman scattering (a detailed analysis will be presented in Section IV). On the other hand, as a single-phonon process the infrared absorption should be barely affected by the anharmonicity.

Since the frequency of $\Gamma_{A\pm}$ depends only on the moiré potential, electronic crystals formed at different fillings in a given triangular moiré pattern then share the same infrared absorption or Raman frequency from $\Gamma_{A\pm}$. Meanwhile the collective nature of the vibration enhances the optical signal by a factor of the total site number. Thus the absorption/Raman peak strength in a triangular crystal at filling 1 is expected to be two times stronger than that in a stripe crystal at filling 1/2. This facilitates the optical measurement of the moiré confinement strength $\gamma$. We note that the optical absorption of $\Gamma_{A\pm}$ is distinct from the single-particle absorption of the Wigner crystal proposed in Ref. [42], which is obtained under a mean-field approximation.

Meanwhile as the inter-site interaction plays no role in $\Gamma_{A\pm}$ modes, the $\Gamma_{A\pm}$-involved infrared absorption and two-photon Raman process should conserve the three-fold rotational ($\hat{C}_3$) symmetry of the underlying moiré potential, even though the lattice configuration of the stripe crystal violates the $\hat{C}_3$-symmetry. For example, absorbing a $\sigma^+$ infrared photon can generate a $\Gamma_{A+}$ phonon with $\sigma^+$ polarization; whereas for an incident $\sigma^+$ photon, the polarizations of the emitted Raman photon and the generated $\Gamma_{A-}$ phonon are both $\sigma^-$. In contrast, the $\Gamma_O$-involved two-photon Raman process is greatly affected by the inter-site interaction, which in the zigzag-stripe crystal doesn't conserve the $\hat{C}_3$ symmetry.

## III. Chiral phonons under time-reversal/inversion symmetry breaking and the phonon Hall effect

In a triangular or inversion-symmetric honeycomb crystal, the doubly degenerate phonons at $\Gamma$ or $K$ can be viewed as the superpositions of opposite circularly-polarized modes (Fig. 3(d)), thus applying an out-of-plane magnetic field will lift the degeneracy and give rise to chiral phonons near these wave vectors. We emphasize that the magnetic field considered here is weak and can be treated as a perturbation, which is distinct from early works where strong magnetic fields are used to localize electrons into Wigner crystals [19,21]. The electronic crystal can also exhibit an out-of-plane ferromagnetic spin order [43], which couples to the circularly polarized vibration through the spin-orbit interaction and plays the same role as an external magnetic field. For the antiferromagnetic spin order in the honeycomb crystal [44], we notice that it violates both the timer-reversal and inversion symmetries, but conserves the parity-time symmetry. Thus the degeneracy at $\Gamma$ and $K$ remains and the chirality is still zero for all non-degenerate modes, which is distinct from the ferromagnetic case.

In triangular crystals, an out-of-plane magnetic field $\mathcal{B}$ interacts with the orbital angular momentums of the electronic sites. When $\mathcal{B} \ll \sqrt{\frac{m\gamma}{\lambda^2}}$ we can keep up to the linear-order of $\mathcal{B}$, the magneto-phonon interaction $\hat{H}_B = \frac{\mathcal{B}}{2m} \cdot \sum_n \mathbf{r}_n \times i\frac{\partial}{\partial \mathbf{r}_n}$ can be expressed with phonon operators as

$$\hat{H}_B = \frac{\mathcal{B}}{2m} \sum_{l,\mathbf{k}} P_{l,\mathbf{k}} \left( \hat{a}_{l,\mathbf{k}}^\dagger \hat{a}_{l,\mathbf{k}} + \frac{1}{2} \right) + \frac{\mathcal{B}}{4m} \sum_{l \neq l',\mathbf{k}} \frac{\omega_{l,\mathbf{k}} + \omega_{l',\mathbf{k}}}{\sqrt{\omega_{l,\mathbf{k}}\omega_{l',\mathbf{k}}}} P_{ll',\mathbf{k}} \hat{a}_{l',\mathbf{k}}^\dagger \hat{a}_{l,\mathbf{k}}$$
$$+ \frac{\mathcal{B}}{8m} \sum_{l \neq l',\mathbf{k}} \frac{\omega_{l,\mathbf{k}} - \omega_{l',\mathbf{k}}}{\sqrt{\omega_{l,\mathbf{k}}\omega_{l',\mathbf{k}}}} P_{ll',\mathbf{k}} \left( \hat{a}_{l,\mathbf{k}} \hat{a}_{l',-\mathbf{k}} - \hat{a}_{l,-\mathbf{k}}^\dagger \hat{a}_{l',\mathbf{k}}^\dagger \right). \tag{9}$$

Here $P_{l,\mathbf{k}} \equiv |A_{l,\mathbf{k}}^+|^2 - |A_{l,\mathbf{k}}^-|^2$ is the circular polarization or chirality of $l$-th phonon branch at $\mathbf{k}$, and $P_{ll',\mathbf{k}} \equiv A_{l,\mathbf{k}}^{+*} A_{l',\mathbf{k}}^+ - A_{l,\mathbf{k}}^{-*} A_{l',\mathbf{k}}^-$ with $l \neq l'$. $\left( A_{l,\mathbf{k}}^+, A_{l,\mathbf{k}}^- \right)$ is the polarization vector of the phonon mode without the magnetic field. Although the small effective mass of the electrons greatly enhances the magneto-phonon interaction as compared to their ionic counterparts, generally the interaction strength is still much weaker than the splitting between different phonon branches ($\mathcal{B}/m \approx 2$ meV for $m = 0.5m_0$ under $\mathcal{B} = 10$ T). Thus the above off-diagonal terms become important only for nearly degenerate modes with $\omega_{l,\mathbf{k}} \approx \omega_{l',\mathbf{k}}$, and the last term can be dropped.

The phonon dispersion and chirality of a triangular crystal under $\mathcal{B}/m = 4$ meV are shown in Fig. 4(a). The originally degenerate modes at $\mathbf{\Gamma}$ and $\mathbf{K}$ are now split into two circularly polarized ones separated by $\mathcal{B}/m$. When slightly displaced from $\mathbf{\Gamma}$ or $\mathbf{K}$, the phonon becomes elliptically polarized whose chirality decays with the displacement. These phonons are then described by the massive Dirac model, which exhibit finite out-of-plane Berry curvatures $\Omega_{l,\mathbf{k}}$. As shown in Fig. 4(b), the Berry curvatures are localized near $\mathbf{\Gamma}$ and $\mathbf{K}$ and can be as large as several hundred to several thousand nm$^2$, meanwhile $\mathbf{K}$ and $\bar{\mathbf{K}}$ have the same Berry curvature due to the inversion symmetry. The Chern numbers of the two phonon branches, however, are both zero. Notice that the chiral phonons near $\mathbf{\Gamma}$ have linear dispersions with finite group velocities, which can be used to transport information encoded on the phonon chirality or magnetic moment [45].

The same analysis can be done to inversion-symmetric honeycomb crystals under a magnetic field, where the chirality is defined as $P_{l,\mathbf{k}} \equiv |A_{l,\mathbf{k}}^+|^2 - |A_{l,\mathbf{k}}^-|^2 + |B_{l,\mathbf{k}}^+|^2 - |B_{l,\mathbf{k}}^-|^2$, and $P_{ll',\mathbf{k}} \equiv A_{l,\mathbf{k}}^{+*} A_{l',\mathbf{k}}^+ + B_{l,\mathbf{k}}^{+*} B_{l',\mathbf{k}}^+ - A_{l,\mathbf{k}}^{-*} A_{l',\mathbf{k}}^- - B_{l,\mathbf{k}}^{-*} B_{l',\mathbf{k}}^-$ for $l \neq l'$. The obtained phonon dispersion and chirality under $\mathcal{B}/m = 4$ meV are shown in Fig. 4(c), with the corresponding Berry curvatures given in Fig. 4(d). The Chern numbers of the four phonon branches are 1, 0, −3 and 2 from low to high frequencies, respectively. When the direction of the magnetic field is reversed, the Berry curvatures and Chern numbers change their signs. Consider two neighboring domains under opposite magnetic field (or ferromagnetic spin) directions, topologically protected edge modes can appear at the interface separating these two domains, which can have potential applications in novel topological phononic devices [46].

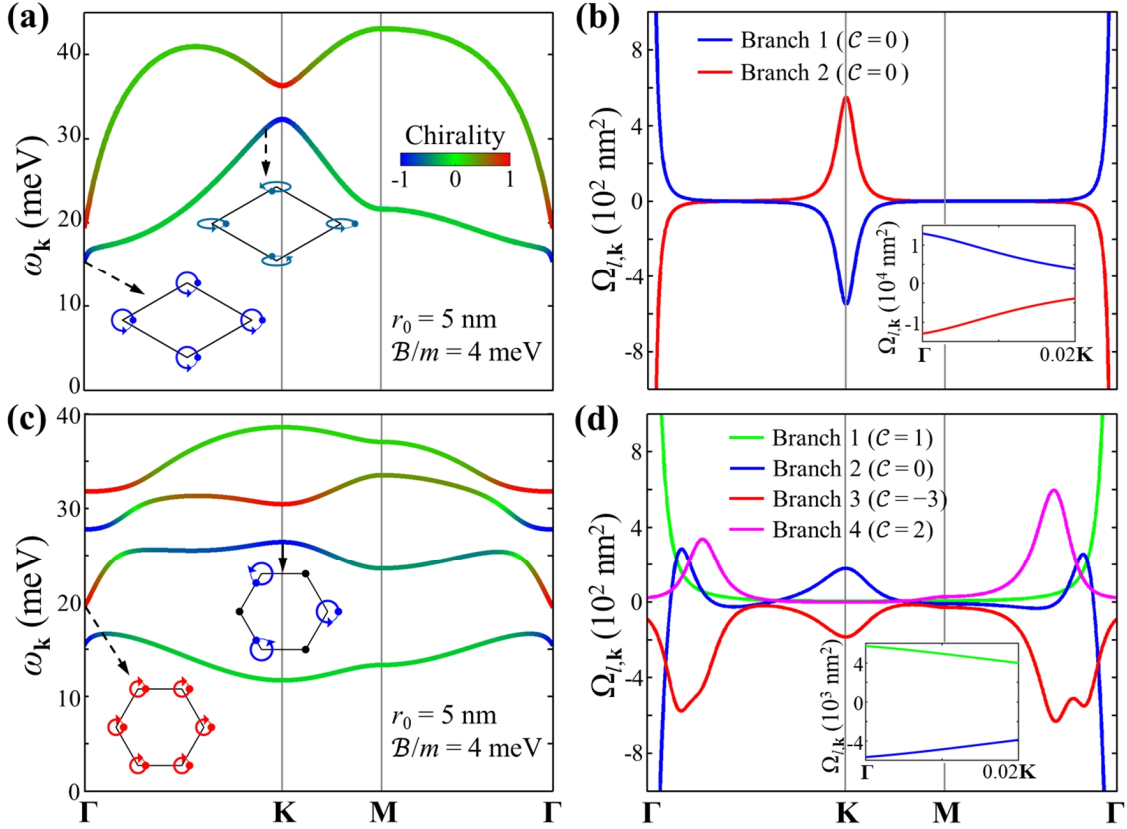

**Figure 4. Chiral phonons in triangular and honeycomb crystals under an out-of-plane magnetic field.** (a) The phonon dispersion in a triangular crystal under an out-of-plane magnetic field $\mathcal{B}/m = 4$ meV. Here $r_0 = 5$ nm, and the other parameters are the same as in Fig. 1(d). The color map shows the chirality of the phonon mode, indicating that phonons at $\mathbf{\Gamma}$ and $\mathbf{K}$ are circularly polarized with chirality $\pm 1$. Phonons slightly away from $\mathbf{\Gamma}$ and $\mathbf{K}$ become elliptically polarized (inset). (b) The calculated Berry curvatures $\Omega_{l,\mathbf{k}}$ for phonons in (a). The inset shows the values of $\Omega_{l,\mathbf{k}}$ near $\mathbf{\Gamma}$. The Chern numbers $\mathcal{C}$ of the two branches are both zero. (c) The phonon dispersion and chirality in an inversion-symmetric honeycomb crystal under an out-of-plane magnetic field $\mathcal{B}/m = 4$ meV. Here $r_0 = 5$ nm, and the other parameters are the same as in Fig. 3(c). (d) The calculated Berry curvatures for phonons in (c), and the Chern numbers $\mathcal{C}$ of the four branches. The inset shows the values of $\Omega_{l,\mathbf{k}}$ near $\mathbf{\Gamma}$ for branch 1 and 2.

In some heterobilayer structures or in homobilayers under an interlayer bias, there can be two inequivalent local minima in each moiré unit cell [30,31,37]. In the strong correlation limit and at a filling factor of 2, both minima will be occupied and the system forms a Mott or charge-transfer insulator with an inversion-asymmetric honeycomb structure. Fig. 5(a) schematically shows an inversion-asymmetric honeycomb crystal and the underlying moiré potential landscape. The calculated phonon dispersion and chirality under the parameters of $m_A = m_B = m_0$, $\gamma_A = 0.2$ eV, $\gamma_B = 0.3$ eV, $\epsilon = 5$, $r_0 = 1$ nm and $\lambda = 10$ nm are given in Fig. 5(b). Such parameters simulate the honeycomb moiré pattern in homobilayer $WS_2$, $MoS_2$, and $MoSe_2$, where the valence band maxima are located at $\Gamma_{TMD}$-valley and $\gamma_B - \gamma_A$ can be tuned by an interlayer bias [33-36]. The $\Gamma_{A\pm}$ and $\Gamma_{O\pm}$ phonons remain doubly degenerate, but both of them become infrared- and Raman-active. The two intermediate modes at **K** are now well-separated and exhibit chirality +1 and −1, respectively. Although calculations show that close to $\Gamma$ the two nearly degenerate optical modes can also exhibit ±1 chirality (Fig. 5(b)), but they are unstable against phonon scattering. The inversion symmetry breaking also gives rise to finite Berry curvatures $\Omega_{l,\mathbf{k}}$ in the vicinity of **K**. The calculated values of $\Omega_{l,\mathbf{K}}$ are shown in Fig. 5(c) as functions of $\gamma_A - \gamma_B$, and $\Omega_{l,\bar{\mathbf{K}}} = -\Omega_{l,\mathbf{K}}$ due to the time-reversal symmetry. As can be seen, the Berry curvature magnitudes of the two chiral phonons ($l = 2,3$) are on the order of several tens to several hundred nm$^2$, and can vary in a large range by tuning $\gamma_B - \gamma_A$ with an interlayer bias. On the other hand, the non-chiral modes ($l = 1,4$) have negligible Berry curvatures.

The large Berry curvatures of chiral phonons in the above time-reversal-asymmetric or inversion-asymmetric crystals can give rise to a phonon Hall effect [25,47,48]. Driven by an in-plane force **F** which can be induced by a spatially inhomogeneous heterostrain [49] or twist angle [50] or interlayer bias (see Appendix D), a chiral phonon with a large Berry curvature $\mathbf{\Omega}_{l,\mathbf{k}}$ acquires an anomalous Hall velocity $\mathbf{\Omega}_{l,\mathbf{k}} \times \mathbf{F}$ which then moves to a transverse direction and accumulate at a sample edge. The circular vibration of the electronic sites gives rise to a finite magnetization proportional to the chirality. The phonon Hall effect then results in a temperature difference between the two edges, as well as a finite magnetization where chiral phonons accumulate (Fig. 5(d)). In crystals under an out-of-plane magnetic field, the equilibrium phonon distribution can already lead to such phenomena since the lowest-frequency $\Gamma_{A\pm}$-phonon is chiral and carries a large Berry curvature (see Fig. 4(a,b)). Whereas in an inversion-asymmetric but time-reversal-symmetric honeycomb crystal, a nonequilibrium distribution will be needed since the Berry curvature and the chirality of the lowest-frequency phonon at **K**/**K̄** are negligible.

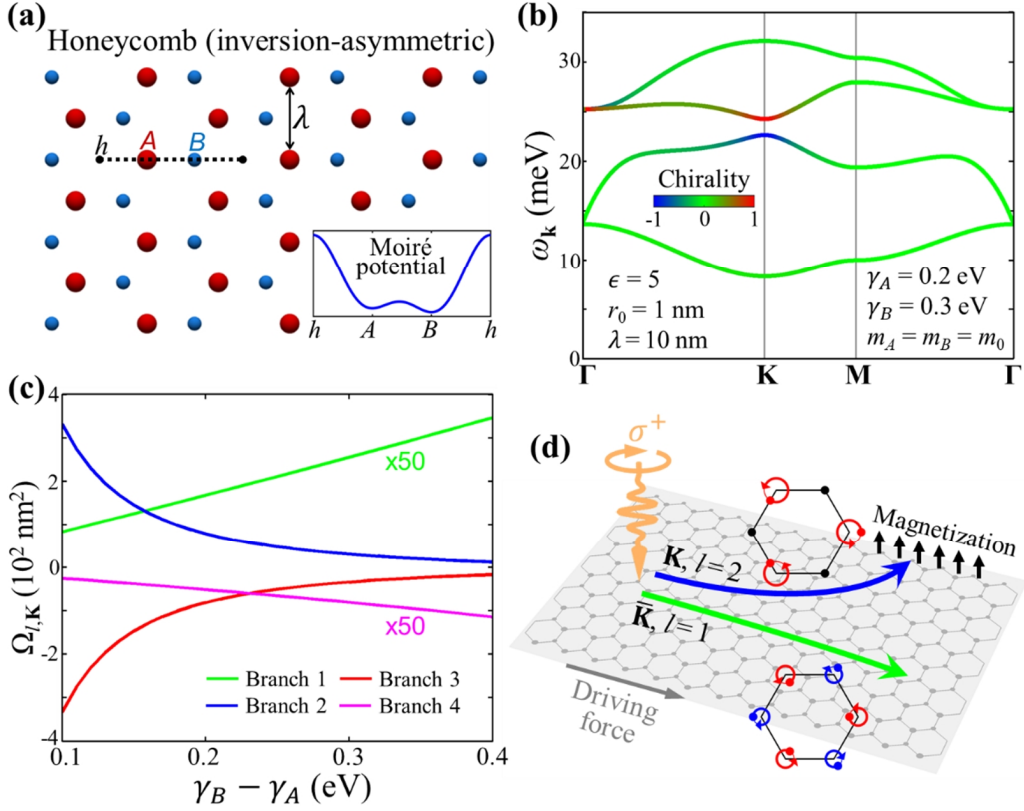

**Figure 5. Chiral phonons in inversion-asymmetric honeycomb crystals.** (a) Schematic illustration of an inversion-asymmetric honeycomb crystal in a honeycomb moiré pattern at a filling factor of 2. The inset shows the moiré potential landscape along the dashed line. (b) The calculated phonon dispersion and chirality under the moiré confinement strengths of $\gamma_A = 0.2$ eV, $\gamma_B = 0.3$ eV. The other parameters are indicated in the figure. (c) The phonon Berry curvatures at **K** as functions of $\gamma_B - \gamma_A$. The other parameters are the same as in (b). Due to the small Berry curvature magnitudes of branch 1 and 4, they have been multiplied by a factor of 50 for better illustration. (d) A schematic illustration of the phonon Hall effect. A $\bar{\mathbf{K}}$-phonon of branch 1 and **K**-phonon of branch 2, whose vibration motions are illustrated, can be excited by absorbing an infrared photon with $\sigma^+$ polarization (see Table I). The **K**-phonon carries a large Berry curvature and chirality −1, which moves to one edge of the sample under the effect of a driving force; whereas the Berry curvature and chirality of the $\bar{\mathbf{K}}$-phonon are negligible. The accumulation of **K**-phonons then results in a finite magnetization at the sample edge.

The **Γ**-phonons can be directly generated through the infrared absorption or two-photon Raman process. Meanwhile, anharmonicities from the moiré and interaction potentials can convert a **Γ**-phonon into two phonons at **K** and $\bar{\mathbf{K}}$, respectively. The combination of these two then allows the optical generation of (**K**, $\bar{\mathbf{K}}$) phonon-pairs. Here we give a qualitative analysis based on the $\hat{C}_3$-symmetry of the triangular and honeycomb crystals, while a quantitative calculation will be presented in Section IV. At the high-symmetry point **Γ** or **K**, the phonon wave function acquires a phase-factor $\exp\left(-i\frac{2\pi}{3}C_3\right)$ upon the in-plane $\hat{C}_3$-rotation about $h$, where $h$ is the empty site of the hexagon formed by $A$ and $B$ sublattices (Fig. 5(a)). $C_3 = 0, \pm 1$ corresponds the pseudo-angular momentum, whose conservation governs the optical selection rule. By properly choosing the polarizations and frequencies of the infrared or Raman photons, specific **Γ**- or (**K**, **K̄**)-phonon modes can be generated with the selection rules summarized in Table I. Taking the process illustrated in Fig. 5(d) as an example, by absorbing a $\sigma^+$ infrared photon with a frequency $\omega_{1,\bar{K}} + \omega_{2,K}$, a **K̄**-phonon in branch 1 and a **K**-phonon in branch 2 can be generated. This **K̄**-phonon (**K**-phonon) has $C_3 = 0$ ($C_3 = -1$) and carries a negligible (large) Berry curvature. Under a driving force, **K**-phonons will move to a transverse direction and accumulate at one edge, whereas the transverse motion of **K̄**-phonons is negligible, resulting in spatially inhomogeneous distributions of the temperature and magnetization.

It should be noted that the $C_3$ quantum number of the **K**/**K̄**-phonon is distinct from its chirality. The chirality gives the average circular polarization of the lattice vibrations and is a local property. On the other hand, the $C_3$ quantum number contains a local contribution which corresponds to the circular polarization, and a nonlocal contribution which accounts for the phase factor $e^{i\mathbf{K} \cdot \mathbf{R}_{A/B,n}}$ in the collective vibration. The nonlocal contribution depends on the choices of the rotation center ($A$, $B$ or $h$), and the $C_3$ quantum numbers shown in Table I are for $h$. We emphasize that for a (**K**, **K̄**)-phonon pair, the sum of their $C_3$ quantum numbers and the selection rules are independent on the choices of the rotation center, as the nonlocal contributions of **K** and **K̄** cancel with each other.

**Table I. The $\hat{C}_3$-symmetry governed selection rules for the optical generation of a Γ-phonon or a (K, K̄)-phonon pair in an inversion-asymmetric honeycomb crystal.** The photon frequency corresponds to the frequency $\omega$ of the infrared photon, or the Raman shift $\Delta\omega$ between the incident and emitted photons. For the infrared absorption, only the $\sigma^+$ photon cases are shown; for the Raman process, only the cases with $\sigma^+$ incident photons are shown. The other cases can be obtained through a time reversal.

| Photon polarization | Photon frequency | Γ-phonon branch | K-phonon branch | K̄-phonon branch |
|---|---|---|---|---|
| $\sigma+$ (infrared absorption) | $\omega = \omega_{l,\Gamma}$ | $\Gamma_{A+}, \Gamma_{O+}$ ($C_3=-1$) | | |
| | $\omega = \omega_{l,K} + \omega_{3,\bar{K}}$ | | $l=1,4$ ($C_3=0$) | $3$ ($C_3=-1$) |
| | $\omega = \omega_{2,K} + \omega_{l,\bar{K}}$ | | $2$ ($C_3=-1$) | $l=1,4$ ($C_3=0$) |
| | $\omega = \omega_{3,K} + \omega_{2,\bar{K}}$ | | $3$ ($C_3=+1$) | $2$ ($C_3=+1$) |
| $\sigma+$ in $\sigma+$ out (Raman process) | $\Delta\omega = \omega_{l,K} + \omega_{l',\bar{K}}$ | | $l=1,4$ ($C_3=0$) | $l'=1,4$ ($C_3=0$) |
| | $\Delta\omega = \omega_{2,K} + \omega_{2,\bar{K}}$ | | $2$ ($C_3=-1$) | $2$ ($C_3=+1$) |
| | $\Delta\omega = \omega_{3,K} + \omega_{3,\bar{K}}$ | | $3$ ($C_3=+1$) | $3$ ($C_3=-1$) |
| $\sigma+$ in $\sigma-$ out (Raman process) | $\Delta\omega = \omega_{l,\Gamma}$ | $\Gamma_{A-}, \Gamma_{O-}$ ($C_3=+1$) | | |
| | $\Delta\omega = \omega_{l,K} + \omega_{2,\bar{K}}$ | | $l=1,4$ ($C_3=0$) | $2$ ($C_3=+1$) |
| | $\Delta\omega = \omega_{3,K} + \omega_{l,\bar{K}}$ | | $3$ ($C_3=+1$) | $l=1,4$ ($C_3=0$) |
| | $\Delta\omega = \omega_{2,K} + \omega_{3,\bar{K}}$ | | $2$ ($C_3=-1$) | $3$ ($C_3=-1$) |

## IV. Anharmonicities and three-phonon processes

When expanding the moiré and interaction potentials about their equilibrium values, the anharmonicity from previous neglected higher-order terms can result in multi-phonon processes. In this section we analyze the effect of leading-order (cubic) anharmonicities in the triangular and honeycomb crystals, and focus on the process that a $\Gamma$-phonon is converted into two phonons with opposite wave vectors. Combined with the optical addressability of $\Gamma$-phonons, this allows the optical generation of phonons at finite wave vectors as illustrated in Fig. 6(a). We shall restrict the following analysis to the initial phonon modes $\Gamma_{A+}$ and $\Gamma_{O+}$, with the anharmonic Hamiltonian given by

$$\hat{H}_{\text{ah}} = \frac{1}{\sqrt{N}} \sum_{l_1 l_2 \mathbf{q}} \left( g_{\Gamma_{A+} \to (l_1 \bar{\mathbf{q}}, l_2 \mathbf{q})} \hat{a}_{\Gamma_{A+}} + g_{\Gamma_{O+} \to (l_1 \bar{\mathbf{q}}, l_2 \mathbf{q})} \hat{a}_{\Gamma_{O+}} \right) \hat{a}^\dagger_{l_1, \bar{\mathbf{q}}} \hat{a}^\dagger_{l_2, \mathbf{q}} + \text{h.c.} \tag{10}$$

Both the moiré potential and the inter-site interaction can contribute to the anharmonicity. The moiré potential at sublattice $\mu = A, B$ can be expanded as $V_{\text{moiré}}(\mathbf{R}_{\mu,n} + \mathbf{r}) \approx V_{\text{moiré}}(\mathbf{R}_{\mu,n}) + \frac{1}{2} \gamma_\mu \frac{r^2}{\lambda^2} + i \beta_\mu \frac{r_+^3 - r_-^3}{\lambda^3}$, with $|\beta_{A,B}| \sim 0.1$ eV from Eq. (2). Its contribution to the three-phonon scattering strength is

$$g^{\text{moiré}}_{\Gamma_{A/O+} \to (l_1 \bar{\mathbf{q}}, l_2 \mathbf{q})} = i (2m\lambda^2)^{-\frac{3}{2}} \frac{\beta_A A^{+*}_{\Gamma_{A/O+}} A^-_{l_1, \bar{\mathbf{q}}} A^-_{l_2, \mathbf{q}} + \beta_B B^{+*}_{\Gamma_{A/O+}} B^-_{l_1, \bar{\mathbf{q}}} B^-_{l_2, \mathbf{q}}}{\sqrt{\omega_{\Gamma_{A/O+}} \omega_{l_1, \mathbf{q}} \omega_{l_2, \mathbf{q}}}}. \tag{11}$$

For the interaction potential, only those between $A$ and $B$ sublattices contributes to the cubic anharmonicity, with a strength given by

$$g^{\text{interation}}_{\Gamma_{A/O+} \to (l_1 \bar{\mathbf{q}}, l_2 \mathbf{q})} = \frac{A^{+*}_{\Gamma_{A/O+}} - B^{+*}_{\Gamma_{A/O+}}}{(2m\lambda^2)^{\frac{3}{2}} \sqrt{\omega_{\Gamma_{A/O+}}}} \left[ \frac{\beta_{0,\mathbf{q}=0} \left( A^-_{l_1, \bar{\mathbf{q}}} A^-_{l_2, \mathbf{q}} + B^-_{l_1, \bar{\mathbf{q}}} B^-_{l_2, \mathbf{q}} \right) - \beta_{0,\mathbf{q}} A^-_{l_1, \bar{\mathbf{q}}} B^-_{l_2, \mathbf{q}} - \beta_{0, \bar{\mathbf{q}}} B^-_{l_1, \bar{\mathbf{q}}} A^-_{l_2, \mathbf{q}}}{\sqrt{\omega_{l_1, \bar{\mathbf{q}}} \omega_{l_2, \mathbf{q}}}} \right.$$

$$\left. - \frac{\beta_{-,\mathbf{q}} \left( A^-_{l_1, \bar{\mathbf{q}}} B^+_{l_2, \mathbf{q}} + A^+_{l_1, \bar{\mathbf{q}}} B^-_{l_2, \mathbf{q}} \right) + \beta_{-, \bar{\mathbf{q}}} \left( B^-_{l_1, \bar{\mathbf{q}}} A^+_{l_2, \mathbf{q}} + B^+_{l_1, \bar{\mathbf{q}}} A^-_{l_2, \mathbf{q}} \right)}{\sqrt{\omega_{l_1, \bar{\mathbf{q}}} \omega_{l_2, \mathbf{q}}}} - \frac{\beta_{+,\mathbf{q}} A^+_{l_1, \bar{\mathbf{q}}} B^+_{l_2, \mathbf{q}} + \beta_{+, \bar{\mathbf{q}}} B^+_{l_1, \bar{\mathbf{q}}} A^+_{l_2, \mathbf{q}}}{\sqrt{\omega_{l_1, \bar{\mathbf{q}}} \omega_{l_2, \mathbf{q}}}} \right], \tag{12}$$

$$\beta_{0,\mathbf{q}} \equiv \frac{\lambda^3}{2\sqrt{2}} \sum_n e^{i\mathbf{q} \cdot (\mathbf{R}_n + \mathbf{R}_{AB})} \partial_-^3 V(\mathbf{R}_n + \mathbf{r})|_{\mathbf{r}=\mathbf{R}_{AB}},$$

$$\beta_{\pm,\mathbf{q}} \equiv \frac{\lambda^3}{2\sqrt{2}} \sum_n e^{i\mathbf{q} \cdot (\mathbf{R}_n + \mathbf{R}_{AB})} \partial_\pm \nabla^2 V(\mathbf{R}_n + \mathbf{r})\big|_{\mathbf{r}=\mathbf{R}_{AB}}.$$

Note that $\beta_{0, \hat{C}_3 \mathbf{q}} = \beta_{0,\mathbf{q}}$, $\beta_{\pm, \hat{C}_3 \mathbf{q}} = e^{\pm 2i\pi/3} \beta_{\mp, \mathbf{q}}$, and $\max(|\beta_{0,q}|) \gg \max(|\beta_{\pm,q}|)$.

In triangular crystals, the cubic anharmonicity from the interaction vanishes. The anharmonicity from the moiré potential plays a critical role for activating the $\Gamma_{A\pm}$-involved Raman process. From Eq. (11), the three-phonon process $\hat{a}_{\Gamma_{A+}} \hat{a}^\dagger_{l_1, \Gamma} \hat{a}^\dagger_{l_2, \Gamma}$ has a strength $g_{\Gamma_{A+} \to (l_1\Gamma, l_2\Gamma)}^{\text{moiré}} \propto A_{\Gamma_{A+}}^{+*} A_{l_1,\Gamma}^{-} A_{l_2,\Gamma}^{-}$, which means a $\Gamma_{A+}$-phonon with the polarization vector $(A_{\Gamma_{A+}}^{+}, A_{\Gamma_{A+}}^{-}) = (1,0)$ can be converted into two $\Gamma_{A-}$-phonons with $(A_{\Gamma_{A-}}^{+}, A_{\Gamma_{A-}}^{-}) = (0,1)$. Considering that a $\Gamma_{A\pm}$-phonon can interconvert with a photon with the $\sigma^{\pm}$ polarization, the Raman process of absorbing a $\sigma^{+}$-photon followed by emitting a $\sigma^{-}$-photon and a $\Gamma_{A-}$-phonon is then activated (see Fig. 6(b)).

Now we consider the strength of the process $\hat{a}_{\Gamma_{A+}} \hat{a}_{l_1,\bar{\mathbf{q}}}^{\dagger} \hat{a}_{l_2,\mathbf{q}}^{\dagger}$ with $\mathbf{q} \neq 0$. Taking the time-reversal-asymmetric triangular crystal of Fig. 4(a) as an example, we show our calculated $|g_{\Gamma_{A+} \to (l_1\bar{\mathbf{q}}, l_2\mathbf{q})}^{\text{moiré}}|^2$ with $(l_1, l_2) = (1,1)$ for all $\mathbf{q}$ values in Fig. 6(c). $(l_1, l_2) = (1,2)$ and $(2,2)$ are found to have much weaker strengths thus are ignored. $\hat{a}_{\Gamma_{A+}} \hat{a}_{1,\Gamma}^{\dagger} \hat{a}_{1,\Gamma}^{\dagger}$ manifests as the most efficient three-phonon process for an initial $\Gamma_{A+}$-phonon. When $\mathbf{q}$ moves away from $\Gamma$ the strength of the three-phonon process decreases sharply, implying that the $\Gamma_{A\pm}$-involved Raman process can be highly efficient.

In honeycomb crystals, the anharmonicity from the interaction (with $|\beta_{\pm,\mathbf{q}}|, |\beta_{0,\mathbf{q}}| \sim 1$ eV) dominates over that from the moiré potential (with $|\beta_{A,B}| \sim 0.1$ eV). Here we consider the process $\hat{a}_{\Gamma_{0+}} \hat{a}_{l_1,\bar{\mathbf{q}}}^{\dagger} \hat{a}_{l_2,\mathbf{q}}^{\dagger}$, whereas switching the initial phonon mode to $\Gamma_{A+}$ only changes the global factor $\frac{A_{\Gamma_{A/0+}}^{+*} - B_{\Gamma_{A/0+}}^{+*}}{\sqrt{\omega_{\Gamma_{A/0+}}}}$ in Eq. (12) but not the dependence on $(l_1\bar{\mathbf{q}}, l_2\mathbf{q})$. In Fig. 6(d) we show the calculated $|g_{\Gamma_{0+} \to (l_1\bar{\mathbf{q}}, l_2\mathbf{q})}^{\text{interation}}|^2$ for the inversion-asymmetric honeycomb crystal of Fig. 5(b). The results are consistent with the symmetry analysis given in Table I. The most efficient processes are found to be $\hat{a}_{\Gamma_{0+}} \hat{a}_{1,\bar{\mathbf{K}}}^{\dagger} \hat{a}_{2,\mathbf{K}}^{\dagger}$, $\hat{a}_{\Gamma_{0+}} \hat{a}_{3,\bar{\mathbf{K}}}^{\dagger} \hat{a}_{1,\mathbf{K}}^{\dagger}$, and $\hat{a}_{\Gamma_{0+}} \hat{a}_{l_1,\Gamma}^{\dagger} \hat{a}_{l_2,\Gamma}^{\dagger}$ with $l_1, l_2 = 3,4$ (note that it's impossible to distinguish branch 3 and 4 near $\Gamma$ as they are almost degenerate). These results indicate that by properly tuning the frequency of the incident photon to match the desired phonon-pair, $(\mathbf{K}, \bar{\mathbf{K}})$-phonon pairs with nonzero chirality and Berry curvatures can be efficiently generated through the infrared absorption or stimulated Raman scattering.

The cubic anharmonicity can slightly renormalize the phonon frequency [51]. However, the chiral phonons at $\Gamma/\mathbf{K}$ are not affected, since the chirality originates from the $\hat{C}_3$-symmetry of the crystal combined with the breaking of the parity-time symmetry, which are not changed by the anharmonicity. The frequency splitting between the ±1 chiral phonons at $\Gamma/\mathbf{K}$ is thus fully determined by the magnetic field $\mathcal{B}$ or the sublattice imbalance $\gamma_B - \gamma_A$, and the chirality near this high-symmetry point should barely change. Meanwhile, the anharmonicity can give rise to a finite linewidth to the phonon. If the frequency splitting between the ±1 chiral phonons is smaller than or comparable to the linewidth, the average chirality and Berry curvature will be near zero. Thus a large enough $\mathcal{B}$ or $\gamma_B - \gamma_A$ is needed to get well separated ±1 chiral phonons. The anharmonicity can also result in a finite thermal conductivity, which quantitatively affects the spatial distributions of the temperature and magnetization in the phonon Hall effect.

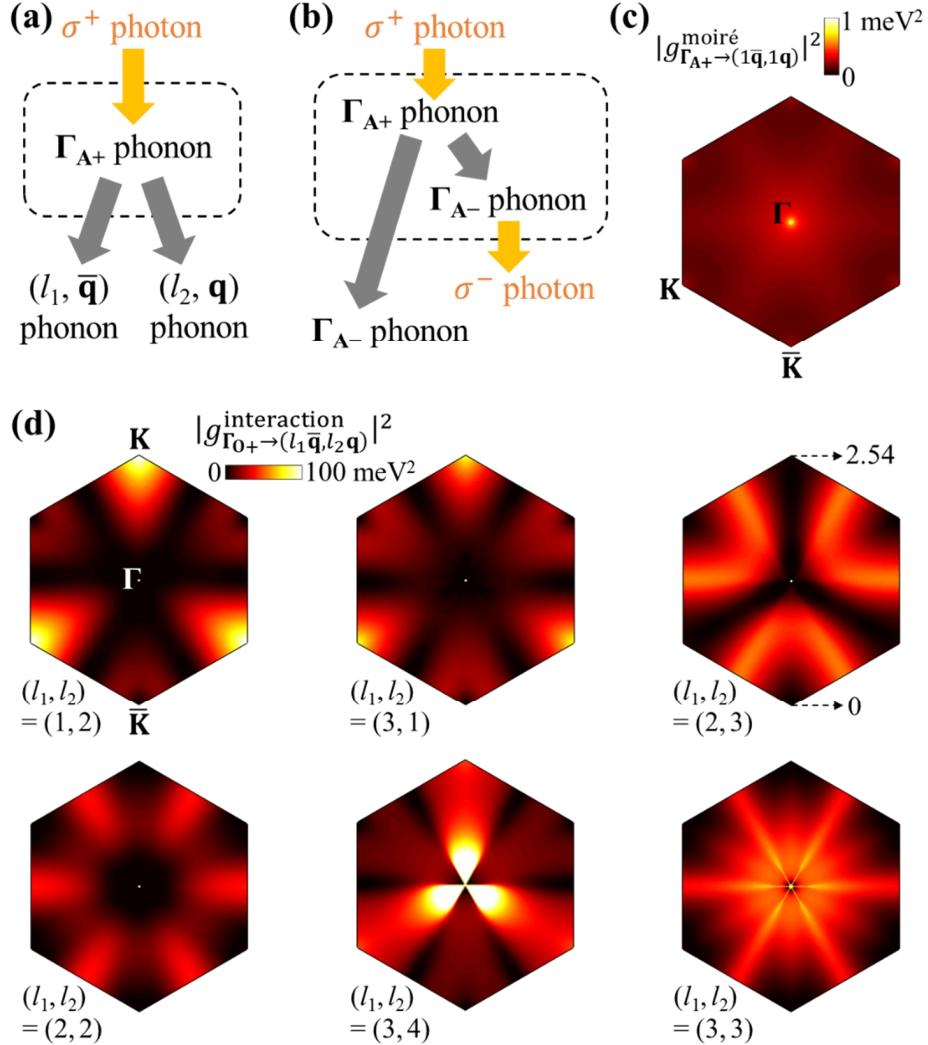

**Figure 6. The anharmonicity induced three-phonon processes.** (a) The optical excitation of a pair of phonons with opposite wave vectors, mediated by a virtual $\Gamma$-phonon. (b) The Raman scattering activated by the three-phonon process $\hat{a}_{\Gamma_{A+}}\hat{a}^\dagger_{l_1,\Gamma}\hat{a}^\dagger_{l_2,\Gamma}$. Those in the dashed boxes are virtual processes. (c) The moiré potential induced scattering strength $\left|g^{\text{moiré}}_{\Gamma_{A+}\to(l_1\bar{\mathbf{q}},l_2\mathbf{q})}\right|^2$ with $(l_1, l_2) =$ **(1,1)** as a function of **q** for the time-reversal-asymmetric triangular crystal of Fig. 4(a). Those with $(l_1, l_2) =$ **(1,2)** and **(2,2)** are found to be much weaker thus not shown. (d) The interaction induced scattering strengths $\left|g^{\text{interaction}}_{\Gamma_{0+}\to(l_1\bar{\mathbf{q}},l_2\mathbf{q})}\right|^2$ as functions of **q** for the inversion-asymmetric honeycomb crystal of Fig. 5(b). Only those $(l_1, l_2)$ with significant scattering strengths are shown. The most efficient three-phonon processes correspond to those **q** near $\mathbf{K}/\bar{\mathbf{K}}$ and $\Gamma$.

## V. Discussion

The emergence of chiral phonons requires either time-reversal or inversion symmetry breaking. Here we briefly discuss possible moiré systems for realizing the inversion-asymmetric honeycomb crystal. In homobilayer WS$_2$, MoS$_2$ and MoSe$_2$ (WSe$_2$ and MoTe$_2$) with a twist angle close to 0, the valence band maxima are located at $\Gamma_{TMD}$ ($\mathbf{K}_{TMD}$) valley [33-36]. In a moiré supercell the potential minima are located at two regions related by a mirror reflection that exchanges the two layers, which then forms an inversion-symmetric honeycomb lattice with degenerate *A* and *B* sublattices. Applying an interlayer bias breaks the sublattice degeneracy, introducing an energy splitting and different moiré confinement strengths $\gamma_A \neq \gamma_B$ between *A* and *B*. In the strong correlation limit and under a filling factor of 2, as long as the on-site repulsion is larger than the energy splitting, the electrons/holes will tend to occupy both sublattices and form an inversion-asymmetric honeycomb crystal. We expect homobilayer WSe$_2$ and MoTe$_2$ to be more suitable platforms, as the valence band maxima at $\mathbf{K}_{TMD}$ are layer-polarized so a small interlayer bias can efficiently break the sublattice degeneracy. On the other hand, the valence band maxima $\Gamma_{TMD}$ in homobilayer WS$_2$, MoS$_2$ and MoSe$_2$ are strongly layer-hybridized, thus a large interlayer bias will be needed to introduce a significant $\gamma_A - \gamma_B$ [34]. Also in homobilayer WSe$_2$, the hole energy at $\Gamma_{TMD}$ is only slightly (~ 50 meV) above that at $\mathbf{K}_{TMD}$ [52]. Under an interlayer bias $\Delta V$, $\mathbf{K}_{TMD}$ at the two sublattices have opposite energy shifts $\pm \Delta V/2$, whereas $\Gamma_{TMD}$ is insensitive to $\Delta V$ [53]. It is then possible to fine-tune $\Delta V$ such that at sublattice *A* (*B*) the valence band maximum is $\mathbf{K}_{TMD}$ ($\Gamma_{TMD}$), which can realize an inversion-asymmetric honeycomb crystal with different effective masses for the two sublattices. Besides the above homobilayer systems, heterobilayer TMDs can also be good candidates. Numerical calculations suggest that some heterobilayer structures have two local potential minima with different energies, which naturally forms an inversion-asymmetric honeycomb superlattice [37].

Our discoveries indicate that electronic crystals in moiré patterns can serve as an exciting new platform to study the chiral phonons, due to their following properties: (1) The moiré system offers various controllability to the electronic crystal. The lattice configuration of the crystal can be controlled by the filling factor, the crystal wavelength can be modulated by the interlayer twist angle, and the Coulomb interaction strength can be modulated through the dielectric patterning of the substrate and cap layer [54-56]. One can also construct a 'two-layer' electronic crystal by vertically stacking two moiré patterns [57], where the filling factors of different moiré patterns can be tuned independently and the phonons are coupled through the interlayer Coulomb interaction [18]. (2) Due to the small effective mass of the crystal sites, phonons couple efficiently to external electric and magnetic fields, as well as the magnetic spin order through the spin-orbit interaction. Chiral phonons with large and tunable Berry curvatures can appear near $\mathbf{\Gamma}$ and $\mathbf{K}$ in triangular/honeycomb crystals, either under the effect of an out-of-plane magnetic field or under an interlayer bias which breaks the inversion symmetry. (3) Chiral phonons at $\mathbf{\Gamma}$ can be directly generated through the infrared absorption or stimulated Raman scattering. Meanwhile, a pair of chiral phonons at $(\mathbf{K}, \overline{\mathbf{K}})$ can be optically generated through the optical addressability of $\mathbf{\Gamma}$-phonons combined with the anharmonicities. These properties facilitate the study of Berry curvatures and the topological effects of phonons, and may have potential applications in novel topological phononic devices. They could also stimulate further theoretical/experimental studies about elementary excitations of the intriguing correlated insulators in moiré systems.

Besides the electronic crystal, the dipole-dipole interaction between interlayer excitons trapped in moiré potential minima can give rise to the excitonic crystal [58,59], and the corresponding signatures have been observed in recent experiments [60-62]. Our analysis to the phonons can be directly applied to that of the excitonic crystal. However, external magnetic and optical fields cannot directly couple to the vibration of the crystal sites because the exciton is charge-neutral. Further studies to effectively implement the coupling between the external fields and phonons of excitonic crystals are needed.

**Acknowledgements:** H.Y. acknowledges support by NSFC under grant No. 12274477, and the Department of Science and Technology of Guangdong Province in China (2019QN01X061).

**Conflict of interest:** The authors declare that there is no conflict of interest.

**Data Availability Statement:** The data that support the findings of this study are available from the corresponding author upon reasonable request.

## Appendix A. The mean-square-displacement and the validity of the harmonic approximation

The mean-square-displacement of the electron site can be expressed as $\langle \mathbf{r}^2 \rangle = \frac{\Omega}{(2\pi)^2 m} \sum_l \int d\mathbf{k} \frac{\langle \hat{a}_{l,\mathbf{k}}^\dagger \hat{a}_{l,\mathbf{k}} \rangle + 1/2}{\omega_{l,\mathbf{k}}}$, where $\langle \hat{a}_{l,\mathbf{k}}^\dagger \hat{a}_{l,\mathbf{k}} \rangle = \frac{1}{e^{\hbar \omega_{l,\mathbf{k}}/k_B T} - 1}$ is the Bose-Einstein distribution. Taking the triangular electronic crystal as an example, we show the contour plots of $\sqrt{\langle \mathbf{r}^2 \rangle}/\lambda$ with the moiré wavelength $\lambda$ and temperature $T$ in Fig. A1(a-c). $\sqrt{\langle \mathbf{r}^2 \rangle}/\lambda$ varies with system parameters, but is generally near 0.2 for our considered wavelength $\lambda \sim 10$ nm. This value is indeed much smaller than 1, implying that our harmonic approximation is valid. According to the Lindemann criterion, the electronic crystal will melt into a quantum liquid when the Lindemann ratio $\sqrt{\langle \mathbf{r}^2 \rangle}/\lambda$ exceeds some critical value ($\approx 0.25$ for the triangular Wigner crystal [63]). We can then use the calculated $\sqrt{\langle \mathbf{r}^2 \rangle}/\lambda$ values to roughly estimate the melting temperature of the electronic crystal. From the data in Fig. A1(a-c), we estimated that the crystal melting occurs at a temperature of several tens Kelvin.

We have also calculated the phonon zero-point energy per site $E_{\text{ZPE}} = \frac{\Omega}{2(2\pi)^2} \sum_l \int d\mathbf{k}\, \omega_{l,\mathbf{k}}$ and the moiré trapping frequency $\lambda^{-1}\sqrt{\gamma/m}$, which are shown in Fig. A1(d). $E_{\text{ZPE}}$ is found to be larger than the moiré trapping frequency, because each electron not only feels the moiré confinement but also the Coulomb repulsion from other electrons. Under the mean-field approximation, the Coulomb repulsion acts as an additional harmonic confinement on the electron site thus increases its zero-point energy.

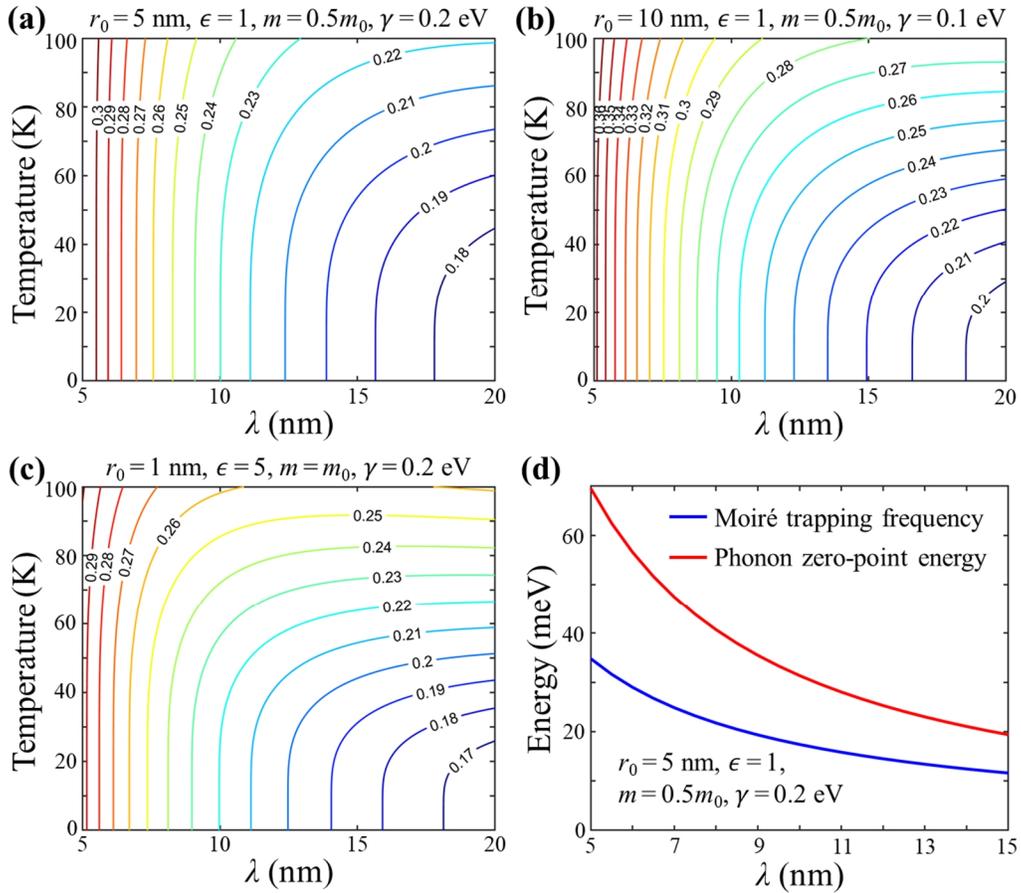

**Figure A1. The mean-square-displacement and the zero-point energies for the triangular electronic crystal.** (a-c) The contour plots of $\sqrt{\langle \mathbf{r}^2 \rangle}/\lambda$ with the moiré wavelength $\lambda$ and temperature $T$, under three

different sets of parameters. (d) The phonon zero-point energy per site and the moiré trapping frequency $\lambda^{-1}\sqrt{\gamma/m}$ as functions of $\lambda$.

## Appendix B. Summation of the Coulomb interaction terms

When evaluating the ground state energy and the phonon dispersion of the crystal, one needs to deal with the summation $\sum_n e^{i\mathbf{k}\cdot\mathbf{R}_n} V(\mathbf{R}_n + \mathbf{r})$. A standard treatment is to directly sum over all lattice vectors $\mathbf{R}_n$ with magnitudes smaller than some cut-off value $R_c$, but the Coulomb potential decay as $r^{-1}$ for large $r$ values thus the summation converges very slowly. To overcome this problem we adopt the 2D Ewald method developed in Ref. [20]. We write the 2D screened Coulomb potential as

$$V(\mathbf{r}) = \frac{\pi}{2r_0}\left[H_0\left(\frac{r}{r_0}\right) - Y_0\left(\frac{r}{r_0}\right)\right] = \int_0^\infty dz \frac{e^{-z}}{\sqrt{r^2 + r_0^2 z^2}}. \qquad (A1)$$

In order to simplify the expression we have drop the factor $\epsilon^{-1}$. Substituting the expression $(r^2 + r_0^2 z^2)^{-\frac{1}{2}} = \frac{1}{\sqrt{\pi}}\int_0^\infty dt \frac{1}{\sqrt{t}} e^{-t(r^2+r_0^2 z^2)}$ into Eq. (A1), we get

$$\sum_n e^{i\mathbf{k}\cdot\mathbf{R}_n} V(\mathbf{R}_n + \mathbf{r}) = \sum_n \frac{e^{i\mathbf{k}\cdot\mathbf{R}_n}}{2r_0} \int_0^\infty \frac{dt}{t} e^{\frac{1}{4r_0^2 t}} \mathrm{erfc}\left(\frac{1}{2r_0\sqrt{t}}\right) e^{-t(\mathbf{R}_n+\mathbf{r})^2}.$$

Here $\mathrm{erfc}(x) = \frac{2}{\sqrt{\pi}}\int_x^\infty dt\, e^{-t^2}$ is the complementary error function. The integral in the above equation can be separated into two parts: $\int_0^\infty = \int_\delta^\infty + \int_0^\delta$, with $\delta$ a separation parameter which will be determined later. The first part can be integrated normally, while the second part is transformed to the momentum space:

$$\frac{1}{2r_0}\int_0^\delta \frac{dt}{t} e^{\frac{1}{4r_0^2 t}} \mathrm{erfc}\left(\frac{1}{2r_0\sqrt{t}}\right) \sum_n e^{i\mathbf{k}\cdot\mathbf{R}_n - t(\mathbf{R}_n+\mathbf{r})^2}$$

$$= \frac{2\pi}{\Omega}\sum_G \frac{e^{-i(\mathbf{k}+\mathbf{G})\cdot\mathbf{r}}}{1 - r_0^2(\mathbf{k}+\mathbf{G})^2}\left[\frac{\mathrm{erfc}\left(\frac{|\mathbf{k}+\mathbf{G}|}{2\sqrt{\delta}}\right)}{|\mathbf{k}+\mathbf{G}|} - r_0 \mathrm{erfc}\left(\frac{1}{2r_0\sqrt{\delta}}\right) e^{\frac{1-r_0^2(\mathbf{k}+\mathbf{G})^2}{4\delta r_0^2}}\right].$$

Here $\mathbf{G}$ is the reciprocal lattice vector of the crystal, and $\Omega$ is the unit cell area. Now we get

$$\sum_n e^{i\mathbf{k}\cdot\mathbf{R}_n} V(\mathbf{R}_n + \mathbf{r}) = \frac{1}{2r_0}\sum_n e^{i\mathbf{k}\cdot\mathbf{R}_n} \int_\delta^\infty \frac{dt}{t}\, \mathrm{erfc}\left(\frac{1}{2r_0\sqrt{t}}\right) e^{\frac{1}{4r_0^2 t} - t(\mathbf{R}_n+\mathbf{r})^2} \qquad (A2)$$

$$+ \frac{2\pi}{\Omega}\sum_G \frac{e^{-i(\mathbf{k}+\mathbf{G})\cdot\mathbf{r}}}{1 - r_0^2(\mathbf{k}+\mathbf{G})^2}\left[\frac{\mathrm{erfc}\left(\frac{|\mathbf{k}+\mathbf{G}|}{2\sqrt{\delta}}\right)}{|\mathbf{k}+\mathbf{G}|} - r_0 \mathrm{erfc}\left(\frac{1}{2r_0\sqrt{\delta}}\right) e^{\frac{1-r_0^2(\mathbf{k}+\mathbf{G})^2}{4\delta r_0^2}}\right].$$

In the above equation, the summations with $\mathbf{R}_n$ and $\mathbf{G}$ both converge rather fast due to the exponential decay. Note that although the result is independent on $\delta$, but a suitable choice of $\delta$ can reduce the cut-off values of $\mathbf{R}_n$ and $\mathbf{G}$ in the summations. In our calculation we set $\delta = \frac{\pi}{\Omega}$, the dynamical matrix and the interaction potential of the electronic crystal can then be calculated.

Special attention should be paid to the static interaction potential per site which has a magnitude $V_0/\mathbf{2}$, here

$$V_0 = \sum_n V(\mathbf{R}_n) - V(\mathbf{r} \to \mathbf{0}) \tag{A3}$$

$$= \sum_{n \neq 0} \int_\delta^\infty \frac{dt}{\mathbf{2}r_0 t} \mathbf{erfc}\left(\frac{t^{-1/2}}{2r_0}\right) e^{\frac{t^{-1}}{4r_0^2} - t\mathbf{R}_n^2} + \frac{2\pi}{\Omega} \sum_{\mathbf{G} \neq 0} \frac{r_0}{1 - r_0^2 \mathbf{G}^2} \left[\frac{\mathbf{erfc}\left(\frac{|\mathbf{G}|}{2\sqrt{\delta}}\right)}{|\mathbf{G}|r_0} - \mathbf{erfc}\left(\frac{1}{2r_0\sqrt{\delta}}\right) e^{\frac{1 - r_0^2 \mathbf{G}^2}{4\delta r_0^2}}\right]$$

$$- \frac{\mathbf{1}}{\mathbf{2}r_0} \int_0^\delta \frac{dt}{t} \mathbf{erfc}\left(\frac{1}{2r_0\sqrt{t}}\right) e^{\frac{1}{4r_0^2 t}} - \frac{\mathbf{2}\pi r_0}{\Omega} \left(e^{\frac{1}{4r_0^2 \delta}} \mathbf{erfc}\left(\frac{1}{2r_0\sqrt{\delta}}\right) - \mathbf{1}\right) - \frac{\mathbf{2}}{\Omega}\sqrt{\frac{\pi}{\delta}} + \frac{\mathbf{1}}{\Omega} \int d\mathbf{r} V(\mathbf{r}).$$

The last term $\frac{1}{\Omega} \int d\mathbf{r} V(\mathbf{r})$ is a singular constant, which vanishes after including the interaction with the neutralizing background charges.

When $r_0 \to \mathbf{0}$, using $\mathbf{erfc}\left(\frac{1}{x}\right)\Big|_{x \to 0} \to x \frac{e^{-1/x^2}}{\sqrt{\pi}}$ we get

$$\sum_n e^{i\mathbf{k} \cdot \mathbf{R}_n} V(\mathbf{R}_n + \mathbf{r})|_{r_0 \to 0} = \sum_n e^{i\mathbf{k} \cdot \mathbf{R}_n} \frac{\mathbf{erfc}(\sqrt{\delta}|\mathbf{R}_n + \mathbf{r}|)}{|\mathbf{R}_n + \mathbf{r}|} + \frac{2\pi}{\Omega} \sum_{\mathbf{G}} e^{-i(\mathbf{k}+\mathbf{G})\cdot\mathbf{r}} \frac{\mathbf{erfc}\left(\frac{|\mathbf{k}+\mathbf{G}|}{2\sqrt{\delta}}\right)}{|\mathbf{k} + \mathbf{G}|},$$

$$V_0|_{r_0 \to 0} = \sum_{n \neq 0} \frac{\mathbf{erfc}(\sqrt{\delta}|\mathbf{R}_n|)}{|\mathbf{R}_n|} + \frac{\mathbf{2}\pi}{\Omega} \sum_{\mathbf{G} \neq 0} \frac{\mathbf{erfc}\left(\frac{|\mathbf{G}|}{2\sqrt{\delta}}\right)}{|\mathbf{G}|} - \mathbf{2}\sqrt{\frac{\delta}{\pi}} - \frac{\mathbf{2}}{\Omega}\sqrt{\frac{\pi}{\delta}} + \frac{\mathbf{1}}{\Omega} \int d\mathbf{r} V(\mathbf{r}).$$

These forms are the same as those obtained in Ref. [20].

## Appendix C. Effects of the site-displacement in zigzag-stripe crystals

The continuously variable site-displacement $\delta x$ plays an important role in the dynamical stability of the zigzag-stripe crystal. When we artificially set $\delta x = 0$ (Fig. A2(a)), the phonon lowest frequency near **M** decreases with the weakening of the screening as indicated in Fig. A2(b), which will eventually cross zero and becomes imaginary when the interaction is strong enough. In contrast, Fig. 2(e) indicates that when $\delta x$ is finite, the lowest frequency barely changes with $r_0$, indicating that the crystal is rather stable even under a strong interaction strength.

Fig. A2(c) indicates the lattice configuration of the zigzag-stripe crystal when $\delta x$ reaches its maximum value $\lambda/4$. The deformed crystal is similar (but not the same) to a triangular crystal which greatly enhances its dynamical stability. Although the zigzag-stripe crystal is unstable when the moiré confinement γ is close to 0, but a weak strength γ ~ 0.02 eV is enough to stabilize it (see Fig. A2(d)). This implies that the zigzag-stripe crystal is highly stable even if the moiré confinement is weak.

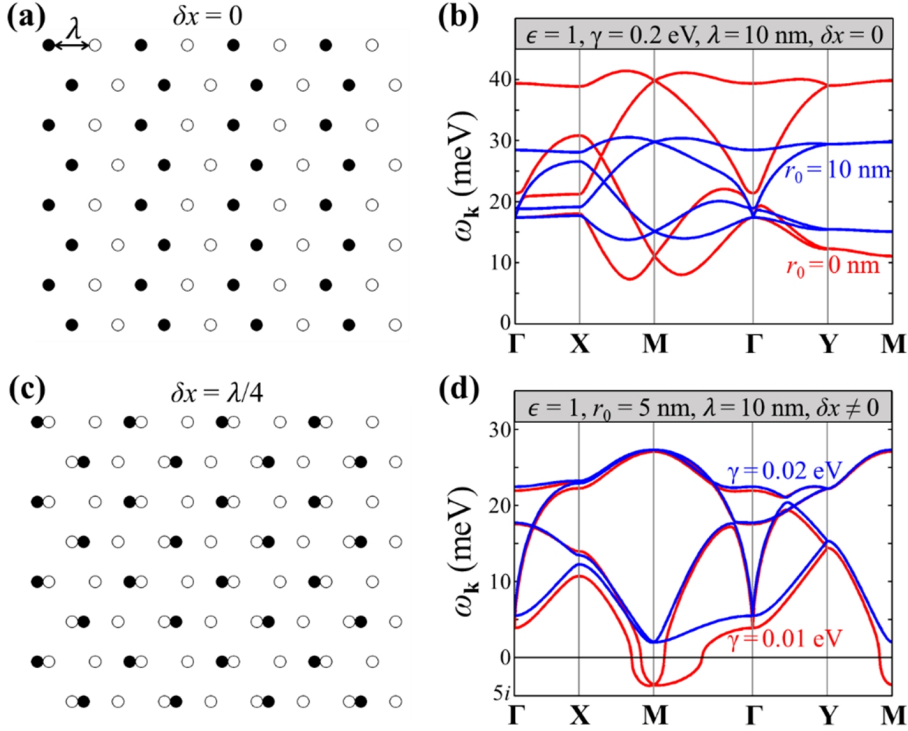

**Figure A2. The effect of site-displacement to phonon dispersions of zigzag-stripe crystals.** (a) The lattice configuration of the zigzag-stripe crystal under $\delta x = 0$. The empty dots correspond to the moiré potential minima, whereas the solid dots are the electronic sites. (b) The phonon dispersions when $\delta x$ is artificially set as 0. Note that the value of the lowest frequency depends sensitively on $r_0$. (c) The lattice configuration when $\delta x$ reaches its maximum value $\lambda/4$. (d) The phonon dispersions under weak moiré confinement strengths when $\delta x$ is finite.

# Appendix D. Phonon driving force induced by the spatially inhomogeneous strain and/or twist angle or interlayer bias

The phonon frequency $\omega_{l,\mathbf{k}}$ is a function of the moiré wavelength $\lambda$. When under the effect of an inhomogeneous heterostrain [49], and/or a continuous varying twist angle which has been realized recently [50], the moiré wavelength $\lambda$ gets modulated with position which gives rise to a driving force $\mathbf{F} = -\frac{\partial \omega_{l,\mathbf{k}}}{\partial \lambda} \nabla \lambda$ applied on the phonon. Notice that the moiré wavelength can be approximated by $\lambda \approx a_U / \sqrt{\delta^2 + \theta^2}$, where $\delta \equiv (a_L - a_U)/a_U$ with $a_U$ and $a_L$ the lattice constants of the upper- and lower-layers, respectively, and $\theta$ is the interlayer twist angle. We assume only $a_L$ is affected by the strain (can be realized when the strain is applied through a substrate), then

$$\frac{\nabla \lambda}{\lambda} \approx -\frac{\delta}{\delta^2 + \theta^2} \frac{\nabla a_L}{a_L} - \frac{\theta}{\delta^2 + \theta^2} \nabla \theta. \quad \text{(A4)}$$

The formation of the moiré pattern requires $\delta, \theta \ll 1$, so generally $\frac{\delta}{\delta^2+\theta^2}, \frac{\theta}{\delta^2+\theta^2} \gg 1$. Eq. (A4) implies that the effect of the inhomogeneous strain (inhomogeneous twist angle) on the moiré wavelength is magnified by a large factor $\frac{\delta}{\delta^2+\theta^2}$ ($\frac{\theta}{\delta^2+\theta^2}$) [64], which can facilitate the application of a phonon driving force.

For the inversion-asymmetric honeycomb crystal in Fig. 5(b), the frequencies of the chiral phonons at $\mathbf{K}/\bar{\mathbf{K}}$ depend on $\gamma_B - \gamma_A$, which can be tuned by an interlayer bias. Thus a spatially inhomogeneous interlayer bias can also apply a driving force on these chiral phonons.